\documentclass[a4paper,12pt]{article}

\usepackage[sumlimits,intlimits,namelimits]{amsmath}
\usepackage{amssymb}
\usepackage{a4wide}
\usepackage{xcolor}
\usepackage{multirow}
\usepackage{nicefrac}
\usepackage{slashed}
\usepackage{comment}
\usepackage{braket}
\usepackage{soul}
\usepackage{float}
\usepackage[normalem]{ulem}

\usepackage[english]{babel}
\makeatletter
\def\fmslash{\@ifnextchar[{\fmsl@sh}{\fmsl@sh[0mu]}}
\def\fmsl@sh[#1]#2{%
  \mathchoice
    {\@fmsl@sh\displaystyle{#1}{#2}}%
    {\@fmsl@sh\textstyle{#1}{#2}}%
    {\@fmsl@sh\scriptstyle{#1}{#2}}%
    {\@fmsl@sh\scriptscriptstyle{#1}{#2}}}
\def\@fmsl@sh#1#2#3{\m@th\ooalign{$\hfil#1\mkern#2/\hfil$\crcr$#1#3$}}
\makeatother
\usepackage{graphicx}
\usepackage{subfigure}
\usepackage{epstopdf}
\usepackage{hep-bibliography}
\usepackage{csquotes}
\usepackage{hyperref}

\bibliography{literature}

\newcommand{\Md}{\mathrm{d}}
\begin{document}
\begin{titlepage}
\begin{flushright}
SI-HEP-2023-29 \\
P3H-23-093 \\
\today
\end{flushright}

\vspace{1.2cm}
\begin{center}
{\Large\bf
Higher twist corrections to $B$-meson decays into a proton and dark antibaryon from QCD light-cone sum rules}
\end{center}

\vspace{0.5cm}
\begin{center}
{\sc Anastasia Boushmelev and Marcel Wald}   \\[2mm]
   Theoretische Physik 1, Center for Particle Physics Siegen \\
   Universit\"at Siegen,  D-57068 Siegen, Germany \\[5mm] 
\end{center}

\vspace{0.8cm}
\begin{abstract}
\vspace{0.2cm}
\noindent
The $B$-Mesogenesis framework anticipates decays of $B$ mesons into a dark antibaryon $\Psi$ and various Standard Model baryons. Here, we focus on the exclusive decay process $B\to p \Psi$ observed as a proton and missing energy in the final state and determine the decay width by employing the QCD light-cone sum rule framework. We include all contributions up to twist six to the nucleon distribution amplitudes in order to parameterize the non-perturbative effects in the operator product expansion. We obtain the decay width and branching fraction with respect to the mass $m_{\Psi}$ of the dark antibaryon $\Psi$, normalized to the model-dependent effective four-fermion coupling.
\end{abstract}

\end{titlepage}

\newpage
\pagenumbering{arabic}
\section{Introduction}
Although the Standard Model of particle physics (SM) is up-to-date the best established theoretical framework to describe particle interactions, it lacks explanations for many observed phenomena like the baryon asymmetry or the dark matter abundance of the universe. Hints on the matter-antimatter asymmetry of the universe can for instance be deduced from measurements of the Cosmic Microwave Background (CMB) \cite{Planck:2015fie,Planck:2018vyg} or the Big Bang Nucleosynthesis (BBN)\cite{Cyburt:2015mya,ParticleDataGroup:2018ovx}.
In general, there exist many different theoretical approaches in the literature which try to address these problems from a theoretical point of view. However, the typical scales involved in these scenarios lie around the Planck scale and are therefore hard to verify experimentally.
The $B$-Mesogenesis model proposed in \cite{Elor:2018twp,Alonso-Alvarez:2021qfd,Alonso-Alvarez:2021oaj,Elahi:2021jia} has emerged as an elegant solution to these puzzles as its features become apparent at measurable energy scales.
\\
Previous studies suggest that the decay $B\to hadrons + \Psi$ is expected to possess appreciable branching fractions with an inclusive width of the order of $10^{-4}$ \cite{Elor:2018twp,Alonso-Alvarez:2021qfd,Alonso-Alvarez:2021oaj}. To explore the feasibility of detecting these modes, a deeper investigation of separate exclusive decay channels becomes important. The original work in \cite{Alonso-Alvarez:2021qfd} roughly estimated the ratios of exclusive to inclusive widths utilizing phase-space counting of quark states. But as it has been shown in \cite{Khodjamirian:2022vta} for the two-particle decay $B \to p \Psi$ and later for different decay channels in \cite{Elor:2022jxy}, the QCD light-cone sum rule (LCSR) approach is well suited to provide estimates for these exclusive decays. \\
LCSRs were initially introduced in \cite{Balitsky:1983sw,Balitsky:1987bk,Chernyak:1990ag} and subsequently applied to various hadronic matrix elements. In \cite{Khodjamirian:2022vta}, the hadronic $B \to p$ transition posed the central challenge in the computation. In order to address this problem, the authors have rather investigated the $p \to B$ transition as it only differs by a global phase to the desired $B \to p$ decay. The advantage of inverting this transition lies in the fact that they need to employ nucleon distribution amplitudes to parameterize the non-perturbative contributions in the LCSR approach, which are studied in greater detail in \cite{Braun:2000kw,Braun:2001tj,Braun:2006hz,Lenz:2009ar,Anikin:2013aka} than $B$-meson distribution amplitudes. In addition to that similar computations have already been carried out in the calculation of form factors for the $\Lambda_b \to p$ transition \cite{Khodjamirian:2011jp}, which are also applicable for the desired $B \to p$ transition. However, only the leading-twist contributions have been considered in order to obtain a first estimate for the corresponding branching fractions. \\
Following the approach from \cite{Khodjamirian:2022vta}, we also focus on the specific two-body decay $B^+ \to p + \Psi$, but include all contributions to the nucleon distribution amplitudes up to twist six. This allows us to perform a dedicated study on the reliability of the leading-twist contributions and the impact of higher twist corrections on the decay width and branching fractions. Moreover, we obtain an estimate on the convergence of the OPE itself.\\
This works is organized as follows: In section \ref{chp:EffOperators}, we introduce the formalism including the basic features of the $B$-Mesogenesis model relevant for the $B \to p$ transition. Therein, we state the input parameters of the model as well as the effective four-fermion operators with the new dark matter particle $\Psi$. Furthermore, section \ref{chp:DerLCSRs} is devoted to the derivation of the LCRSs, while section \ref{chp:CorrFunc} illustrates the computation of the various OPE contributions. This leads to the expressions for the various form factors, which we extrapolate to the physical timelike region in section \ref{chp:zExp}. Section \ref{chp:Numerics} introduces the parameters of the nucleon distribution amplitude and their dependence on the renormalization scale $\mu$ and subsequently shows the numerical evaluation of these form factor expressions in subsection \ref{chp:FormFactors} and of the branching fractions in subsection \ref{chp:BranchingFractions}. Finally, we conclude in section \ref{chp:Conclusion}. Appendix \ref{chp:NucleonDA} and \ref{chp:AppendixFormFactors} provide further supplementary information.

\section{Effective operators} \label{chp:EffOperators}

\begin{figure}[h]
	\centerline{
		\includegraphics[width=7cm, height=3cm]{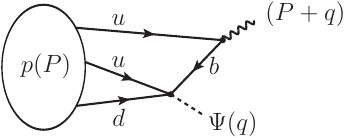}}
	\caption{Diagram for the $p \to B$ transition taken from \cite{Khodjamirian:2022vta}, which differs from the $B \to p$ transition describing the $B \to p \Psi$ decay by an unobservable global phase. The mediator particle $Y$ has been integrated out such that we obtain an effective four-fermion interaction containing the new dark matter particle $\Psi$.}
	\label{fig:DiagramBMesogenesis}
\end{figure}

As we have already pointed out, the decay $B \to p \Psi$ is one of the simplest decay channels which lead to the introduction of a dark sector restoring baryon number conservation by considering the combination of the SM and the dark sector. In order for the dark matter particle $\Psi$ to be observable and not to decay into SM particles immediately, it is only allowed to interact gravitationally with the SM or via a heavy colour-triplet scalar field $Y$ with a mass of the order of a few TeV. However, the hypercharge of the scalar mediator particle $Y$ is not unique, there exist in general two different possibilities with $Q_Y=-1/3$ and $Q_Y=2/3$. Throughout this work, we only focus the model with $Q_Y=-1/3$ for simplicity, because these considerations can be similarly applied to the second case. \\
The part of the Lagrangian governing the additional interactions of the $Y$-field with SM quarks and the dark matter field $\Psi$ is given by
\begin{align}
	{\cal L}_{(Q_Y = -1/3)}=&-y_{ud}\epsilon_{ijk}Y^{*\,i}\bar{u}_R^{\,j}d_R^{c\,k}
	-y_{ub}\epsilon_{ijk}Y^{*i}\bar{u}_R^{\,j}b_R^{c\,k}
	-y_{\Psi d}Y_i\bar{\Psi} d_R^{c\,i}  -y_{\Psi b}Y_i\bar{\Psi} b_R^{c\,i} + \mathrm{h.c.} \;  \label{eq:FullLagrangian}
\end{align}
with $c(R)$ denoting charge conjugated (right-handed) fields, $q_R = \frac12 (1+\gamma_5)q$, while $i, j, k$ indicate the colour indices of the quarks in fundamental representation. Notice that we expect a completely antisymmetric combination of colour charged fields in the interaction of $Y$ with SM quarks in order to preserve gauge invariance \cite{Elor:2018twp}. Additionally, the quantities $y_{ud}, y_{\Psi d}$ and $y_{ub}, y_{\Psi b}$ represent the (antisymmetric) Yukawa couplings between the dark sector and the SM sector.\\
Similar to \cite{Khodjamirian:2022vta}, we exploit that the mass of the interaction particle $Y$, $M_Y$, is much greater compared to the typical momentum transfers of this decay $k \sim m_B$ and therefore integrate out the heavy mediator $Y$. Effectively, the propagator turns into
\begin{align}
	\frac{i}{k^2 - M_Y^2} = -\frac{i}{M_Y^2} \cdot \Big(1 + \frac{k^2}{M_Y^2} + ...\Big) \approx -\frac{i}{M_Y^2}
\end{align}
such that we obtain the following effective Lagrangian including four-fermion interactions
\begin{align}
	{\cal L}_{(Q_Y = -1/3)}\, =& \,
 	\frac{y_{ub}y_{\Psi d}}{M_Y^2} i \epsilon_{ijk}\left(\bar{\Psi}d_R^{c\,i}\right)\left(\bar{u}_R^{j} b_R^{c\,k}\right)
	+  \frac{y^*_{ub}y^*_{\Psi d}}{M_Y^2} i \epsilon_{ijk}\left(\bar{b}^{c\,i}_R u^{\,j}_R\right)\left(\bar{d}^{c\,k}_R\Psi\right) +
	\{d\leftrightarrow b \}\, .
	\label{eq:Leff1}
\end{align}
We depict the corresponding Feynman diagram for the effective interaction in figure \ref{fig:DiagramBMesogenesis}.
 \noindent
From the expression in Eq. \eqref{eq:Leff1}, we can immediately read off the effective Hamiltonian
\begin{align}
	{\cal H}_{(Q_Y = -1/3)}\, =& \,
	- \frac{y_{ub}y_{\Psi d}}{M_Y^2} i \epsilon_{ijk}\left(\bar{\Psi}d_R^{c\,i}\right)\left(\bar{u}_R^{j} b_R^{c\,k}\right)
	- \frac{y^*_{ub}y^*_{\Psi d}}{M_Y^2} i \epsilon_{ijk}\left(\bar{b}^{c\,i}_R u^{\,j}_R\right)\left(\bar{d}^{c\,k}_R\Psi\right)+
	\{d\leftrightarrow b \}\,.
	\label{eq:Heff1}
\end{align}
For the extraction of the effective three-quark operators from the four-fermion interaction in Eq. \eqref{eq:Heff1}, it is useful to employ the Fierz identity \cite{Nieves:2003in}
\begin{align}
	\bar{\Psi}d_R^c=\bar{d}_R \Psi^c, ~~\bar{d}_R^c\Psi=\bar{\Psi}^cd_R \, .
	\label{eq:Fierz}
\end{align}
After that we factorize the field $\Psi$ from the effective four-fermion interaction and obtain
 \begin{align}
	{\cal H}_{(Q_Y = -1/3)} \, =
	- \; G_{(d)}\bar{{\cal O}}_{(d)} \Psi^c 
	-\:
	G^*_{(d)}\bar{\Psi}^c{\cal O}_{(d)}+
	\{d\leftrightarrow b \}\,~~~~
	\label{eq:Heff2}
\end{align}
with the effective four-fermion coupling $G_{(d)} = (y_{ub}y_{\Psi d})/M_Y^2$ and define the local three-quark operator and its conjugate
\begin{align}
	\bar{{\cal O}}_{(d)}= i \epsilon_{ijk}\left(\bar{u}^i_{R} b_R^{c\,j}\right)\bar{d}^k_R, ~~
	{\cal O}_{(d)}= i \epsilon_{ijk}d^{\,i}_R \left(\bar{b}^{c\,j}_R u^{\,k}_R\right) .
	\label{eq:Od}
\end{align}
It is also possible for the $b$-quark to couple to the dark matter particle $\Psi$, which leads to the following operators:
\begin{align}
	\bar{{\cal O}}_{(b)}= i \epsilon_{ijk}\left(\bar{u}^i_{R} d_R^{\,c\,j}\right)\bar{b}^k_R, ~~
	{\cal O}_{(b)}= i \epsilon_{ijk}b^{\,i}_R \left(\bar{d}^{\,c\,j}_R u^{\,k}_R\right) \; .
	\label{eq:Ob}
\end{align}
In the remaining analysis, we consider the operators in Eqs. \eqref{eq:Od} and \eqref{eq:Ob} as two individual versions of the $B$-Mesogenesis model and call them $(d)$- and $(b)$-model, respectively. This is in analogy to \cite{Alonso-Alvarez:2021qfd}, where these operators are referred to as "type-II" and "type-I" operators. These operators in Eq. \eqref{eq:Od}, \eqref{eq:Ob} constitute the central elements of our framework, since the correlation function includes these operators for the determination of the $B \to p \Psi$ decay.
The decay amplitude for this particular decay in the $(d)$-version of the $B$-Mesogenesis model is given by 
\newpage
\begin{align}
	\mathcal{A}_{(d)}(B^+ \to p \Psi) =& \; G_{(d)} \bra{p(P) \Psi^c} \bar{\cal{O}}_{(d)} \ket{B^+(P + q)} =  G_{(d)} \bra{p(P)} \bar{\mathcal{O}}_{(d)} \ket{B^+(P + q)} u_{\Psi}^c(q)\,
	\label{eq:AmplitudeStart}
\end{align}
and similarly for the $(b)$-model if we replace the operator in Eq. \eqref{eq:AmplitudeStart} by the corresponding operator from Eq. \eqref{eq:Ob}. We choose the momentum assignment according to figure \ref{fig:DiagramBMesogenesis} such that the on-shell conditions read $(P + q)^2 = m_B^2,\, p^2 = m_p^2$ and $q^2 = m_{\Psi}^2$. Furthermore, we decompose the $B \to p$ transition into the four different form factors
\begin{align}
	\langle p(P)| \bar{{\cal  O}}_{(d)} |B^+(P+q)\rangle =
	F^{(d)}_{B\to p_R}(q^2)\bar{u}_{p,R}(P)+
	F^{(d)}_{B\to p_L}(q^2)\bar{u}_{p,L}(P)
	\nonumber\\
	+
	\widetilde{F}^{(d)}_{B\to p_R}(q^2)\bar{u}_{p,R}(P) \frac{\slashed{q}}{m_p}+ \widetilde{F}^{(d)}_{B\to p_L}(q^2)\bar{u}_{p,L}(P) \frac{\slashed{q}}{m_p} \; .
	\label{eq:ffDecomp}
\end{align}
Note that we introduce the factor $1/m_p$ for the last two Dirac structures to render the form factors dimensionless. After replacing $(d) \to (b)$, we obtain the corresponding form factors for the $(b)$-model. These set of form factors will be determined via the light-cone sum rule approach in the following sections.

\section{Derivation of the light-cone sum rules}  \label{chp:DerLCSRs}

The correlation function plays the key role in the derivation of the light-cone sum rules, since it directly connects the physical timelike region with the perturbatively calculable spacelike region via the quark-hadron duality (QHD). Therefore, we begin this discussion by stating the form of the correlation function which corresponds to the effective framework introduced in the last section \ref{chp:EffOperators}
\begin{align}
	\Pi^{(d)}(P,q)=i\int  \mathrm{d}^4x\ e^{i(P+q)\cdot
		x}\bra{0}T\left\{j_B(x),{\cal O}_{(d)} (0)\right\}\ket{p(P)} , 
	\label{eq:CorrFuncd}
\end{align}
where $j_B(x) = i m_b \bar{b}(x) \gamma_5 u(x)$ is the $B$-meson current. For the second version of the $B$-mesogenesis model, the operator $\mathcal{O}_{(d)}$ needs to be replaced by $\mathcal{O}_{(b)}$ from Eq. \eqref{eq:Ob}. Contrary to the discussion above, we investigate the $p \to B$ transition rather than the necessary $B \to p$ transition for the decay $B \to p \Psi$. These transitions differ at most by a global phase which do not alter physical observables like decay widths or branching fractions. One advantage is that we can parameterize the long-distance contributions in terms of nucleon light-cone distribution amplitudes, which are better known than the $B$-meson DAs \cite{Braun:2000kw,Braun:2001tj,RQCD:2019hps,Anikin:2013aka,Braun:2006hz,Lenz:2009ar}. Especially the parameters of the nucleon DAs have been determined to better accuracy by advanced lattice computations and sum rules analyses such that we can perform our analysis to twist six accuracy. In this context, the second advantage is that we can closely follow the analysis for $\Lambda_b \to p$ form factors from \cite{Khodjamirian:2011jp}. Although the currents inside the correlation function differ for this problem, the computation of the form factors requires to use the same distribution amplitudes. We state the decomposition of the nucleon matrix element, its transformation into distribution amplitudes of definite twist and the relevant shape parameters in appendix \ref{chp:NucleonDA}. \\
The next step is to make use of the unitarity condition and the Schwartz reflection principle. While the former corresponds to the insertion of a complete set of states inside Eq. \eqref{eq:CorrFuncd}, the latter expresses this correlation function in terms of a dispersion relation in $(P+q)^2$. As it is usually the case for transitions involving $B$-mesons, we can easily separate the ground state contribution in form of a $B$-meson pole. Beyond the threshold cutoff $s_h = (m_B + 2m_{\pi})^2$, we observe hadronic contributions like excited states and continuum contributions, which we incorporate into the hadronic spectral density $\rho^{h(d)}$. 
Employing in the hadronic dispersion relation in $(P+q)^2$, we can access the form factors via
\begin{align}
	\Pi^{(d)}(P,q)=\frac{\langle 0 | j_B|B^+(P+q)\rangle
		\langle B^+(P+q)| {\cal  O}_{(d)} |p(P)\rangle
	}{m_B^2-(P+q)^2}+
	\int\limits_{s_h} ^\infty ds\frac{\rho^{h(d)}(s,P,q)}{s-(P+q)^2}\,.
	\label{eq:HadronicDispersion}
\end{align}
Notice that we ignore possible subtraction terms in this context as they vanish during a Borel transformation, which additionally suppresses the continuum contributions and leads to a better convergence of the sum rules.
The correlation function in Eq. \eqref{eq:CorrFuncd} consists of different kinematical contributions, which are given by all possible Lorentz-invariant amplitudes:
\begin{align}
	\Pi^{(d)}(P,q)=\Pi^{(d)}_R((P+q)^2,q^2) u_{p,R}(P)+\Pi^{(d)}_L((P+q)^2,q^2) u_{p,L}(P)
	\nonumber\\
	+\widetilde{\Pi}^{(d)}_R((P+q)^2,q^2)\slashed{q} u_{p,R}(P)+
	\widetilde{\Pi}_L^{(d)}((P+q)^2,q^2)\slashed{q} u_{p,L}(P)\,.
	\label{eq:CorrFuncDecomp}
\end{align}
Here, we use that the Dirac spinors $u_{p,\{R,L\}}(P)$ are the right-handed and left-handed components of the Dirac spinor $u_p(P)$, i.e. $u_{p,\{R,L\}}(P) = P_{\{R,L\}} u_p(P) = \frac12 (1 \pm \gamma_5) u_p(P)$. Furthermore, contributions involving structures like $\slashed{P} u_{p,\{R,L\}}$ can be included in the first line of Eq. \eqref{eq:CorrFuncDecomp} due to the Dirac equation.\\
Based on the decomposition into different kinematical contributions, we can derive individual dispersion relations for the various form factors in Eq. \eqref{eq:ffDecomp}.
For this, we insert \eqref{eq:CorrFuncDecomp} into Eq. \eqref{eq:HadronicDispersion} and group contributions according to their different Dirac structures such that we can directly access the individual form factors. For example, the dispersion relation for the form factor $F^{(d)}_{B\to p_R}(q^2)$ reads
\begin{align}
\Pi^{(d)}_R((P+q)^2,q^2)=\frac{m_B^2f_B\,F^{(d)}_{B\to p_R}(q^2)}{m_B^2-(P+q)^2}+ \;
	\!\!\!\int\limits_{s_h} ^\infty\!ds\frac{\rho_R^{h(d)}(s,q^2)}{s-(P+q)^2}\,,
	\label{eq:HadronicDispersionR}
\end{align}
where we exploit that $\bra{0}j_B\ket{B^+}=m_B^2f_B$. In addition to that we decompose the hadronic spectral density $\rho^{h(d)}$ in terms of the decomposition in Eq. \eqref{eq:CorrFuncDecomp}. The other form factors can be obtains similarly by the following replacements:
\begin{align}
&&
\Pi^{(d)}_R\to& \,\Pi^{(d)}_L,~\widetilde{\Pi}^{(d)}_R,~\widetilde{\Pi}^{(d)}_L,  
~~~F^{(d)}_{B\to p_R} \to F^{(d)}_{B\to p_L}, ~m_p^{-1}\widetilde{F}^{(d)}_{B\to p_R},~
m_p^{-1} \widetilde{F}^{(d)}_{B\to p_L}\,,
\nonumber\\
&&
\rho_R^{h(d)}\to& \rho_L^{h(d)},~ \widetilde{\rho}_R^{\,h(d)},~ \widetilde{\rho}_L^{\,h(d)}\,. 
\label{eq:repl}
\end{align}
Note that we introduce a factor of $1/m_p$ in front of the form factors $F^{(d)}_{B\to p_{R,L}}(q^2)$ to obtain the correct mass dimensions.
Before heading to the calculation of the correlation function, which will be covered in the next section, we need to find a proper expression for the hadronic spectral density. All results from the next section can be expressed via a dispersion relation of the OPE contributions, which we compute in the deep spacelike region by employing perturbative methods. For the amplitude $\Pi_R^{(d)}$, it takes the form
\begin{align}
	\Pi^{(d), \text{OPE}}_R((P+q)^2,q^2)=\frac{1}{\pi}\int\limits_{m_b^2} ^\infty ds\frac{\mbox{Im}\Pi_R^{(d),\text{OPE}}(s,q^2)}{s-(P+q)^2}\,,
	\label{eq:OPEdsip}
\end{align}	
where we write $\rho_R^{(d),\text{OPE}} = \frac{1}{\pi} \mathrm{Im} \Pi_R^{(d),\text{OPE}}$. In order to remove the hadronic spectral density, which is hard to describe from a theoretical point of view due to its complicated structure, we replace the integral over the hadronic spectral density $\rho^{h(d)}$ by an integral over the spectral density obtained from the OPE $\frac{1}{\pi}\mathrm{Im} \Pi^{(d),\text{OPE}}$ using the (semi-local) quark-hadron duality
\begin{align}
	\int\limits_{s_h} ^\infty ds\frac{\rho_R^{h(d)}(s,q^2)}{s -(P+q)^2}=
	\frac{1}{\pi}\int\limits_{s_0^B} ^\infty ds\frac{\mbox{Im}\Pi_R^{(d),\text{OPE}}(s,q^2)}{s -(P+q)^2}
	\,.
	\label{eq:DualityLCSRs}
\end{align}
This step introduces the effective threshold $s_0^B$, which needs to be determined in the numerical analysis as it is an input parameter in the sum rule framework. \\
Finally, we substitute Eq. \eqref{eq:OPEdsip} into Eq. \eqref{eq:HadronicDispersionR} and perform a Borel transformation in the variable $(P + q)^2$ to arrive at the desired form of the sum rules:
\begin{align}
m_B^2f_B\,F^{(d)}_{B\to p_R}(q^2)e^{-m_B^2/M^2}=
\frac{1}{\pi}\int\limits_{m_b^2} ^{s_0^B} ds\, e^{-s/M^2}\mbox{Im}\Pi_R^{(d),\text{OPE}}(s,q^2)\,.
\label{eq:LCSR} 
\end{align}
As we have previously discussed, we obtain the other form factors with the replacements from Eq. \eqref{eq:repl}.

\section{Correlation Function}  \label{chp:CorrFunc}
Now that we derived the form factors in terms of the sum rules corresponding to Eq. \eqref{eq:LCSR}, we can evaluate the OPE to the desired twist accuracy and to leading order in $\alpha_s$. 
\begin{align}
\Pi^{(d)}(P,q)&=-i \varepsilon^{ijk}m_b\int d^4 x e^{i(P+q)\cdot x}\bra{0}T\{[\overline{b}(x)\gamma_5u(x)],d_R^i(0) [(u_R^j(0))^TCb_R^k(0)]\}\ket{p(P)}
\label{eq:correlationfct}
\end{align}
In order to apply light-cone sum rules, we need to work in the phase space region where $(P + q)^2 \ll m_b^2$ and $q^2 \ll m_b^2$ is valid, which means that the momenta of the involved particles are far off-shell. According to figure \ref{fig:DiagramBMesogenesis}, the $b$-quarks inside the correlation function in Eq. \eqref{eq:correlationfct} are connected to form a $b$-quark propagator. This leaves us with a proton-to-vacuum matrix element containing two uncontracted $u$-quark fields as well as a $d$-quark, which encodes the non-perturbative information in the $p \to B$  transition. Following the usual procedure of the light-cone sum rule approach, we replace this matrix element by nucleon DAs by decomposing the matrix element in terms of different Lorentz structures based on Lorentz invariance and parity invariance. We obtain in total 24 structures if we consider all contributions up to twist six accuracy, which can be subsequently related to distribution amplitudes of definite twist. We provide all details regarding this procedure and the further parameterization of the shape of the distribution amplitudes in the conformal expansion in appendix \ref{chp:NucleonDA}. \\
Contrary to previous investigations \cite{Khodjamirian:2022vta}, see also \cite{Elor:2022jxy}, we consider all contributions up to twist six accuracy including the $\mathcal{O}(x^2)$ corrections to the leading-twist contributions. $\mathcal{O}(x^2)$ corrections to twist four contributions, which correspond to a twist six effect, are numerically negligible \cite{Braun:2001tj} and not considered in this work. By including these set of contributions, we are able to estimate the reliability of the leading-twist analyses from \cite{Khodjamirian:2022vta} since we can study the convergence of the OPE and observe the impact of higher twist corrections on the branching fraction of the $B \to p \Psi$ decay itself.
We start the computation by reproducing the known leading-twist results from \cite{Khodjamirian:2022vta}:
\begin{align}
	\Pi^{(d)}(P,q)=& \; \Bigg[- \frac{ m_b}{2}\int\limits_0^1 \Md\alpha
	\frac{\big((1-\alpha)m_p^2+P\cdot q\big)(V_1 + A_1)(\alpha)}{((1-\alpha)P+q)^2-m_b^2}\Bigg]u_{p,R}(P)
	\label{eq:PidTwist3} \\
	\Pi^{(b)}(P,q)=& \; \Bigg[-\frac{ m_b m_p}{4}\int\limits_0^1 \Md\alpha
	\frac{(1-\alpha)m_p(V_1 + A_1)(\alpha)-3m_b T_1(\alpha)}{((1-\alpha)P+q)^2-m_b^2}\Bigg]u_{p,R}(P)
	\nonumber
	\\
	&-\Bigg[\frac{ m_bm_p}{4}\int\limits_0^1 \Md\alpha\frac{(V_1 + A_1)(\alpha)}{((1-\alpha)P+q)^2-m_b^2} 
	\Bigg]\slashed{q}u_{p,L}(P)\,.
	\label{eq:PibTwist3}
\end{align}
In general, the nucleon DAs introduce three different variables $\alpha_{1,2,3}$ representing the momentum fractions of the individual quarks inside the proton. In Eqs. \eqref{eq:PidTwist3} and \eqref{eq:PibTwist3}, we have integrated over $\alpha_2$ and $\alpha_3$ and renamed $\alpha = \alpha_1$. Moreover, we observe that for the $(d)$-model only one form factor contributes, while there is an additional contribution for the $(b)$-model from the Lorentz structure $\slashed{q} u_{p,L}(P)$. We state the leading-twist distribution amplitudes $V_1, A_1, T_1$ in appendix \ref{chp:NucleonDA}, where we also elaborate on the other higher twist distribution amplitudes in more detail. \\
Scalar products of the form $P \cdot q$ are not suitable for Borel transformations, hence we replace them by
\begin{align}
	2 P \cdot q = (P + q)^2 - m_p^2 - q^2 \, .
\end{align}
This allows us to cancel the $(P + q)^2$-dependence in the denominators of Eqs. \eqref{eq:PidTwist3} and \eqref{eq:PibTwist3} after rewriting
\begin{align}
	((1-\alpha)P + q)^2 = (1 - \alpha) (P + q)^2 + \alpha q^2 - \alpha (1 - \alpha) m_p^2 \;. \label{eq:DenominatorRewr}
\end{align}

\begin{table}[h!]
	\centering
	{
		\scalebox{1.0}{
		\begin{tabular}{| c || c |} 
			\hline
			&\\[-3.5mm]
			twist $3$ & $\int \mathrm{d}^d x \; e^{ikx} e^{i(P + q)x} e^{-iP \alpha x} = (2 \pi)^d \delta^{(d)}(k + q + P \bar{\alpha})$ \\
			&\\[-3.5mm]
			\hline
			&\\[-3.5mm]
			\multirow{4.25}{*}{twist $4,5,6$} & $\int \mathrm{d}^d x \; e^{ikx} e^{i(P + q)x} e^{-iP \alpha x} x_{\nu} = -i (2 \pi)^d \frac{\partial}{{\partial k^{\nu}}} \delta^{(d)}(k + q + P \bar{\alpha})$ \\
			&\\[-3.5mm]
			\cline{2 -- 3}
			&\\[-3.5mm]
			& $\int \mathrm{d}^d x \; e^{ikx} e^{i(P + q)x} e^{-iP \alpha x} x_{\mu} x_{\nu} = (-i)^2 (2 \pi)^d \frac{\partial}{{\partial k^{\mu}}} \frac{\partial}{{\partial k^{\nu}}} \delta^{(d)}(k + q + P \bar{\alpha})$ \\
			&\\[-3.5mm]
			\cline{2 -- 3}
			&\\[-3.5mm]
			& $\int \mathrm{d}^d x \; e^{ikx} e^{i(P + q)x} e^{-iP \alpha x} x_{\mu} x^{\mu} = (-i)^2 (2 \pi)^d \frac{\partial}{{\partial k^{\mu}}} \frac{\partial}{{\partial k_{\mu}}} \delta^{(d)}(k + q + P \bar{\alpha})$ \\[1mm]
			\hline
		\end{tabular}}}
	\caption{Integration over the position variable $x$ with additional factors of $x_{\mu}$ starting at twist four accuracy. For brevity, we introduce the notation $\bar{\alpha} = 1 - \alpha$.}
	\label{tab:MomConservation}
\end{table} \noindent
Constant terms independent of $(P + q)^2$ vanish under the subsequent Borel transformation. \\
Higher twist corrections become more involved, because they explicitly show $x$-dependencies in the expressions. First of all, they occur as explicit factors $x_{\mu}$ in our calculation, which we rewrite according to table \ref{tab:MomConservation} as derivatives acting on the momentum-conserving $\delta$-distribution. 
Besides that we deal with scalar products of the form $P \cdot x$, which need to be introduced in order to relate the 24 invariant functions $\mathcal{S}_i, \mathcal{P}_i, \mathcal{A}_i, \mathcal{V}_i, \mathcal{T}_i$ in the decomposition in Eq. \eqref{eq:DecomNuclDAs} to distribution amplitudes of definite twist, see appendix \ref{chp:NucleonDA}, and for instance \cite{Braun:2000kw,Braun:2001tj} for more details. An additional partial integration with respect to the variable $\alpha$ removes these factors and we additionally notice that the occurring surface terms vanish. \\
With these steps in mind, we can perform a similar derivation of the form factors as in \cite{Khodjamirian:2022vta}. We intend to bring the contributions from Eqs. \eqref{eq:PidTwist3} and \eqref{eq:PibTwist3}, including now all contributions up to twist six, into the form of an dispersion integral.
\noindent
For this, we perform the following substitution after using Eq. \eqref{eq:DenominatorRewr}
\begin{align}
	s := \frac{m_b^2 - \alpha q^2 + \alpha (1 - \alpha) m_p^2}{1 - \alpha} \, . \label{eq:Substitution}
\end{align}
Finally, we need to perform the Borel transformation to obtain the final form of the LCSRs for the form factors
\begin{align}
\mathcal{B}_{M^2}\frac{1}{(s+Q^2)^k}=\frac{1}{(k-1)!}\left(\frac{1}{M^2}\right)^{k-1}e^{-s/M^2} \, ,
\end{align}
where $\mathcal{B}_{M^2}$ denotes the Borel transformation. In our case we identify $Q^2=-(P+q)^2$ and we encounter the cases $k=1,2,3$. This is related to the fact that the denominators in Eqs. \eqref{eq:PidTwist3} and \eqref{eq:PibTwist3} occur to higher powers if we consider the higher twist corrections. Ultimately, this leads to a suppression of these factors in powers of $1/M^2$. Performing all these steps for instance for the $F^{(d)}_{B\to p_R}(q^2)$ form factor results into\\
\begin{align}
	F^{(d)}_{B\to p_R}(q^2) =& \, \frac{1}{m_B^2 f_B} \int_0^{\alpha_0^B} \Md \alpha \, e^{\frac{m_B^2 - s(\alpha)}{M^2}} \Bigg\{\frac{m_b^3}{4} \bigg(1 + \frac{\bar{\alpha}^2 m_p^2 - q^2}{m_b^2}\bigg) \frac{(V_1 + A_1)(\alpha)}{\bar{\alpha}^2} \nonumber \\ &- \frac{m_b^2 m_p}{2} \frac{P_1(\alpha)+S_1(\alpha)}{\bar{\alpha}} + \frac{m_b m_p^2}{4}\bigg( V_3(\alpha)-A_3(\alpha)\bigg)   \nonumber \\ &+ \frac{m_b^3 m_p^2}{4M^2} \frac{\widetilde{V}_{123}(\alpha)-\widetilde{A}_{123}(\alpha)}{\bar{\alpha}^2} \bigg(1 + \frac{m_p^2 \bar{\alpha}^2 - q^2}{m_b^2}\bigg)  + \frac{m_b^2 m_p^3}{2} \frac{\widetilde{S}_{12}(\alpha)-\widetilde{P}_{21}(\alpha)}{\bar{\alpha}M^2} \nonumber \\ &- \frac{m_b m_p^2}{4} \frac{\widetilde{V}_{1345}(\alpha)+\widetilde{A}_{1345}(\alpha)}{\bar{\alpha}} \bigg(1 + \frac{m_b^2}{\bar{\alpha} M^2}\bigg)  + \frac{m_b m_p^2}{4 \bar{\alpha}^2}\bigg(\widetilde{A}_1^M-\widetilde{V}_1^M\bigg) \nonumber \\ &\times \bigg(1 + \frac{q^2 - m_p^2 \bar{\alpha}^2 + m_b^2}{\bar{\alpha}M^2} + \frac{m_b^2}{\bar{\alpha}^2 M^4} \Big(q^2 - m_p^2 \bar{\alpha}^2 - m_b^2\Big)\bigg)  \nonumber \\ &+ \frac{m_b m_p^4}{2} \frac{\widetilde{\widetilde{V}}_{123456}(\alpha)-\widetilde{\widetilde{A}}_{123456}(\alpha)}{\bar{\alpha}M^2} \bigg(1 + \frac{m_b^2}{\bar{\alpha}M^2}\bigg) \Bigg\} \, . \label{eq:FdBpR}
\end{align}
We state the relevant notation and various functions in appendix \ref{chp:NucleonDA}. Notice that we introduce the notation 
\begin{align}
	\widetilde{V}(\alpha) =& \, \int_0^{\alpha} \Md \alpha' \, V(\alpha') \\
	\widetilde{\widetilde{V}}(\alpha) =& \; \int_{0}^{\alpha} \Md \alpha' \int_{0}^{\alpha'} \Md \alpha'' \, V(\alpha'') \, ,
\end{align}
to denote the distribution amplitudes which we integrate in $\alpha$ once or twice in order to remove the scalar product $P \cdot x$. With the replacements from Eq. \eqref{eq:repl}, we can derive similar expressions for the other form factors. It turns out that the $(d)$-model receives additional contributions for the $\widetilde{F}^{(d)}_{B\to p_L}(q^2)$ form factor such that we end up with two different form factors in each model. However, we note that the $\mathcal{T}$-structures in the proton matrix element in the decomposition \eqref{eq:DecomNuclDAs} result into sizeable effects for the $(b)$-model, while they vanish in the $(d)$-model. We provide the remaining form factor in the $(d)$-model and the form factors for the $(b)$-model in appendix \ref{chp:AppendixFormFactors}.

\section{Extrapolation to the large $m_{\Psi}$-region} \label{chp:zExp}

In the last section, we have obtained the form factors for both the $(d)$- and $(b)$-model, which ultimately enter the decay width for the process $B \to p \Psi$. These form factors are valid in the limit $q^2 \ll m_b^2$ since we have used the light-cone sum rule approach, meaning that we are still working on the light-cone with small distances. However, the kinematics of this two-particle decay shows that the upper bound of the particle mass $m_{\Psi}$ is given by $m_B - m_p \approx 4.34$ GeV, while the lower bound lies around $m_p$ in order to prevent proton decays \cite{Elor:2018twp,Alonso-Alvarez:2021oaj,Alonso-Alvarez:2021qfd}. Hence, the mass of the dark matter particle $\Psi$ is theoretically allowed to be in the range $m_{\Psi} \sim m_b$. \\

In order to extract reliable results from the form factors expressions derived in the last section, we perform an extrapolation of Eqs. \eqref{eq:FdBpR} and \eqref{eq:formdpl} to \eqref{eq:formbpl} using the BCL-version \cite{Bourrely:2008za} of the $z$-expansion \cite{Boyd:1995cf}. For this, we perform a conformal mapping of the variable $q^2$ onto the complex variable $z$:

\begin{align}
    z(q^2)=(\sqrt{t_+-q^2}-\sqrt{t_+-t_0})/(\sqrt{t_+-q^2}+\sqrt{t_+-t_0})
\end{align}
with $t_0 = (m_B+m_p)\cdot (\sqrt{m_B}-\sqrt{m_p})^2$ and $t_\pm=m_B\pm m_p$. While $t_0$ is a default parameter, usually chosen as above, in order to minimize the truncation error of the $z$-expansion, $t_\pm$ are set by the physics of the decay. The parameter $t_- = m_B - m_p$ is precisely the upper bound on $m_\Psi$ dictated by the two-particle decay of the $B$ meson at rest and $t_+ = m_B + m_p$ constitutes the threshold for multiparticle states and higher resonances. Starting from this threshold, the timelike form factors become imaginary, which is also represented by the variable $z$ developing an imaginary part. However, the choice of $t_+$ introduces another subtlety, namely an isolated pole due to the $\Lambda_b$-baryon at $q^2 = m_{\Lambda_b}^2$. The treatment of these particular issues is further described in \cite{Bourrely:2008za} such that we finally end up with 
\begin{align}
	F^{(d)}_{B \to p_R}(q^2) =& \; \frac{F^{(d)}_{B \to p_R}(0)}{1 - q^2/m_{\Lambda_b}^2} \Bigg[1 + b^{(d)}_{B \to p_R} \bigg(z(q^2) - z(0) + \frac{1}{2} \Big[z(q^2)^2 - z(0)^2\Big]\bigg)\Bigg]\, . \label{eq:FFzExpFinal}
\end{align}

Here, it is sufficient to truncate the $z$ expansion to $\mathcal{O}(z^2)$, because the allowed range for the mass $m_\Psi$ results in small values of $z$ in the interval $0.077 > z > -0.083$. This leaves us with two free parameters which we need to determine further. These two parameters are given by the form factor evaluated at $q^2=0$, which we can determine directly from Eqs. \eqref{eq:FdBpR} and \eqref{eq:formdpl} to \eqref{eq:formbpl}, and the slope parameter $b^{(d)}_{B \to p_R}$, which we get from the fitting procedure in the next section. For the other form factors, one just has to perform the replacement rule in Eq. \eqref{eq:repl} and additionally $(d)\to (b)$.

\section{Numerical analysis} \label{chp:Numerics}

\begin{table}[h]
	\centering
	\scalebox{0.85}{
	\begin{tabular}{|c||c|c|}
		\hline
		&&\\[-3.0mm]
		Parameter  &  interval & Ref.  \\
		&&\\[-3.0mm]
		\hline
		\hline
		&&\\[-3.0mm]
		$b$-quark $\overline{\text{MS}}$ mass & $\overline{m}_b(3~\mbox{GeV})= 4.47^{+0.04}_{-0.03} $ GeV 
		& \cite{ParticleDataGroup:2020ssz}
		\\
		&&\\[-3.0mm]
		\hline
		&&\\[-3.0mm]
		Renormalization scale &$\mu =3.0^{+1.5}_{-0.5}$ GeV&
		\multirow{3.25}{*}{\cite{Khodjamirian:2017fxg,Khodjamirian:2020mlb}}
		\\[1mm]
		Borel parameter squared &  $M^2=16.0\pm 4.0 $ GeV$^2$ &\\[1mm]
		Duality threshold  & $ s_0^B =39.0^{-1.0}_{+1.5} $ GeV$^2$ &  \\
		&&\\[-3.0mm]
		\hline
		&&\\[-3.0mm]
		$B$-meson decay constant & $f_B= 190.0\pm 1.3$ MeV & \cite{FlavourLatticeAveragingGroupFLAG:2021npn}
		\\
		&&\\[-3.0mm]
		\hline
		&&\\[-3.0mm]
		Nucleon decay constant &  $f_N(\mu=2~\mbox{GeV}) = \big(3.54^{+0.06}_{-0.04}\big)\times 10^{-3}~ \mbox{GeV}^2$ 
		&
		\cite{RQCD:2019hps}\\
		&&\\[-3.0mm]
		\hline
		&&\\[-3.0mm]
		& $\varphi_{10}(\mu=2~\mbox{GeV})=0.182^{+0.021}_{-0.015}$&\\[1mm]
		& $\varphi_{11}(\mu=2~\mbox{GeV})=0.118^{+0.024}_{-0.023}$&\multirow{2.5}{*}{\cite{RQCD:2019hps}}
		\\[1mm]
		& $\lambda_{1}(\mu=2~\mbox{GeV}) = \big(-44.9^{+4.2}_{-4.1}\big)\times 10^{-3}~ \mbox{GeV}^2$&\\[1mm]
		Parameters of nucleon DAs & $\lambda_{2}(\mu=2~\mbox{GeV}) = \big(93.4^{+4.8}_{-4.8}\big)\times 10^{-3}~ \mbox{GeV}^2$&\\[1mm]
		& $\eta_{10}(\mu=\sqrt{2}~\mbox{GeV})=-0.039^{+0.005}_{-0.005}$&\multirow{2.5}{*}{\cite{Anikin:2013aka}}
		\\[1mm]
		& $\eta_{11}(\mu=\sqrt{2}~\mbox{GeV})=0.14^{+0.016}_{-0.016}$&\\[1mm]
		& $\xi_{10}(\mu=2~\mbox{GeV}) = -0.042^{+0.313}_{-0.312}$& \cite{Braun:2006hz}
		\\[1mm]
		\hline
	\end{tabular}}
	\caption{Input parameters in the LCSRs from the references in this Table.}
	\label{tab:input}
\end{table} \noindent
The input parameters for the LCSRs are given in table \ref{tab:input}. We perform our analysis at a renormalization scale of $\mu = 3$ GeV. This particular choice is in accordance with recent studies on the $B \to \pi$ transition or $B^*B \pi$ strong couplings \cite{Khodjamirian:2017fxg,Khodjamirian:2020mlb}, because these analyses indicate that this scale and its uncertainty is optimally suited for $B$-meson interpolating currents. Additionally, we adopt the $b$-quark mass in the $\overline{\text{MS}}$-scheme and use the $B$-meson decay constant obtained from recent lattice QCD computations with $n_f = 2+1+1$ \cite{FlavourLatticeAveragingGroupFLAG:2021npn}.

As table \ref{tab:input} shows, the input parameters for the nucleon distribution amplitude are extracted from different sources. For example, a recent lattice calculation \cite{RQCD:2019hps} determines some input parameters at the scale $\mu_0 = 2$ GeV, whereas a LCSR computation \cite{Anikin:2013aka} works at $\mu_0 = \sqrt{2}$ GeV. Therefore, we make use of the following RGE to run these parameters to required scale $\mu = 3$ GeV  
\begin{align}
	\frac{\text{d}}{\text{d} \ln \mu} \; \varphi(\mu) = -\gamma_{\varphi} \; \varphi(\mu) \;,
\end{align}
with $\gamma_{\varphi}$ being the non-cusp anomalous dimension for the DA parameter $\varphi$.
Its solution to one-loop order is given by
\begin{align}
	\varphi(\mu) = \varphi(\mu_0) \, \bigg(\frac{\ln(\mu_0/\Lambda_{\text{QCD}})}{\ln(\mu/\Lambda_{\text{QCD}})}\bigg)^{\frac{\gamma_{\varphi}^0}{2 \beta_0}} \; , \label{eq:RGESolPhiFinal}
\end{align}
where $\gamma_{\varphi}^0$ denotes the one-loop anomalous dimension. 
The value for $\Lambda_{\text{QCD}} = 0.288$ GeV is taken from \cite{Herren:2017osy} for $n_f = 4$.
Analogously, Eq. \eqref{eq:RGESolPhiFinal} can be used in order to run the remaining nucleon DA parameters to the desired scale $\mu = 3$ GeV. This requires the one-loop non-cusp anomalous dimensions $\gamma_{\varphi}^0$ for all parameters, which we give in table \ref{tab:AnomDim} following \cite{Anikin:2013aka}.

\begin{table}[h]
	\centering
	\begin{tabular}{|c||c|c|c|c|c|c|c|c|}
		\hline
		&&&&&&&&\\[-4mm]
		Parameter $\varphi$  & $f_N$ &  $\varphi_{10}$ & $\varphi_{11}$ & $\eta_{10}$ & $\eta_{11}$ & $\lambda_{1}$ & $\lambda_{2}$ & $\xi_{10}$ \\
		&&&&&&&&\\[-4mm]
		\hline
		\hline
		&&&&&&&&\\[-4mm]
		$\gamma_{\varphi}^0$ & $\frac43$ & $\frac{40}{9}$ & $\frac{16}{3}$ & $\frac{40}{9}$  & $8$ & $4$ & $4$ & $\frac{20}{3}$ \\[1mm]
		\hline
	\end{tabular}
	\caption{One-loop non-cusp anomalous dimensions for different parameters of the nucleon DA in table \ref{tab:input}.}
	\label{tab:AnomDim}
\end{table}\noindent
These parameters can be related to the parameters of the conformal expansion via \cite{Anikin:2013aka}
\begin{align}
	A_1^u =& \; \varphi_{10} + \varphi_{11} \, , \nonumber \\
	V_1^d =& \; \frac13 - \varphi_{10} + \frac13 \varphi_{11} \, , \nonumber \\
	f_1^d =& \; \frac{3}{10} - \frac16 \frac{f_N}{\lambda_1} + \frac15 \eta_{10} - \frac13 \eta_{11} \, , \nonumber \\
	f_1^u =& \; \frac{1}{10} - \frac16 \frac{f_N}{\lambda_1} - \frac35 \eta_{10} - \frac13 \eta_{11} \, , \nonumber \\
	f_2^d =& \; \frac{4}{15} + \frac25 \xi_{10} \nonumber \; .
\end{align}

\subsection{Form factors}  \label{chp:FormFactors}
\begin{figure}[H]
	\centerline{
		\includegraphics[width=14cm, height=7cm]{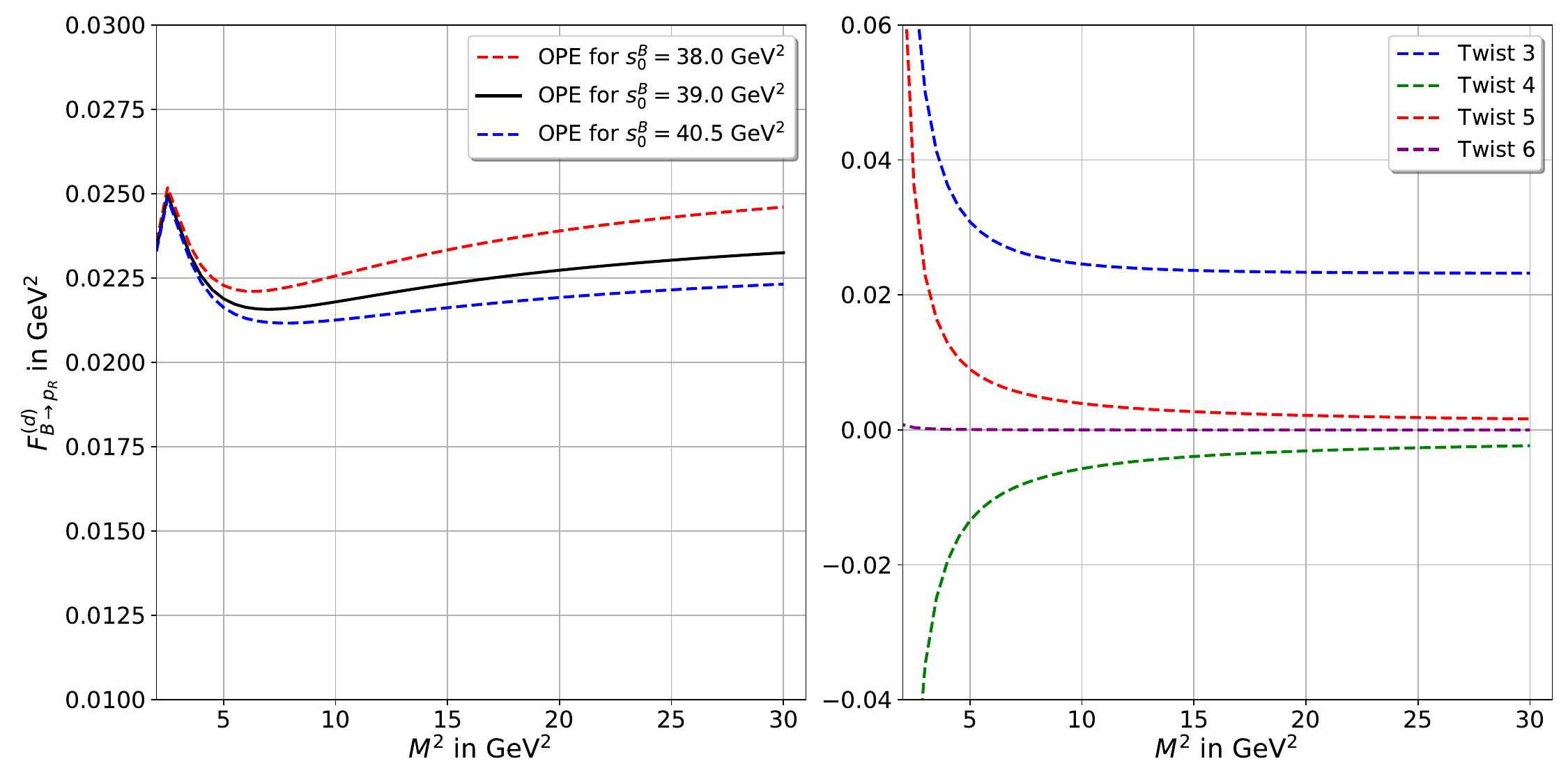}
	}
	\caption{Analysis for the form factor $F_{B \to p_R}^{(d)}(q^2)$ in the $(d)$-model for different Borel parameter $M^2$. The left plot shows all twist contributions combined for various choices of the threshold parameter $s_0^B$, while the right plot illustrates each twist individually. The mass of the dark matter particle $m_{\Psi}$ is set to the benchmark value $m_{\Psi} = 2$ GeV \cite{Alonso-Alvarez:2021qfd}.}
	\label{fig:FBpRd}
\end{figure}

\begin{figure}[H]
	\centerline{
		\includegraphics[width=14cm, height=7cm]{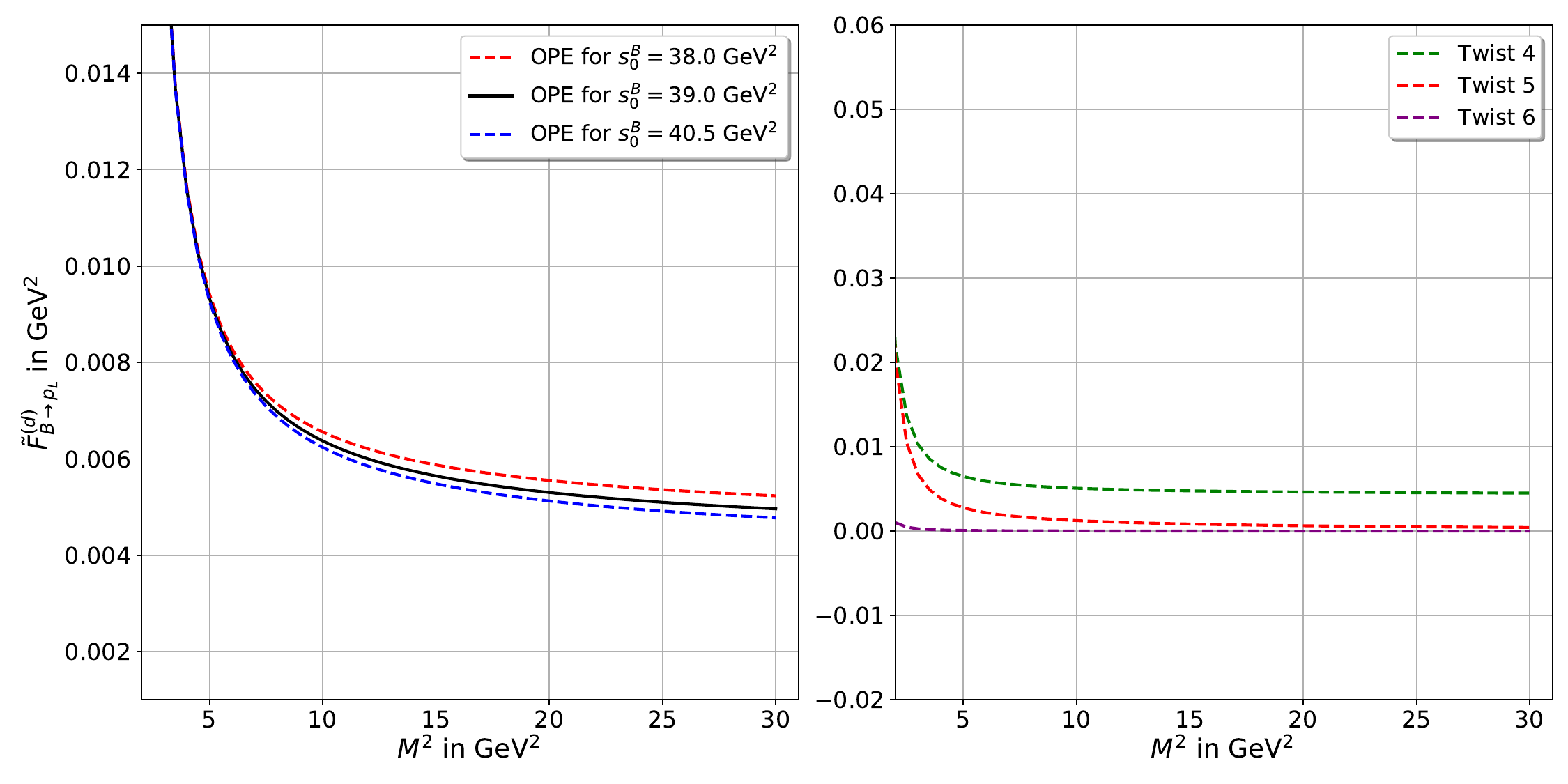}
	}
	\caption{Similar analysis as in figure \ref{fig:FBpRd} for the form factor $\widetilde{F}_{B \to p_L}^{(d)}(q^2)$.}
	\label{fig:FBpLd}
\end{figure}

\begin{figure}[h]
	\centerline{
		\includegraphics[width=14cm, height=7cm]{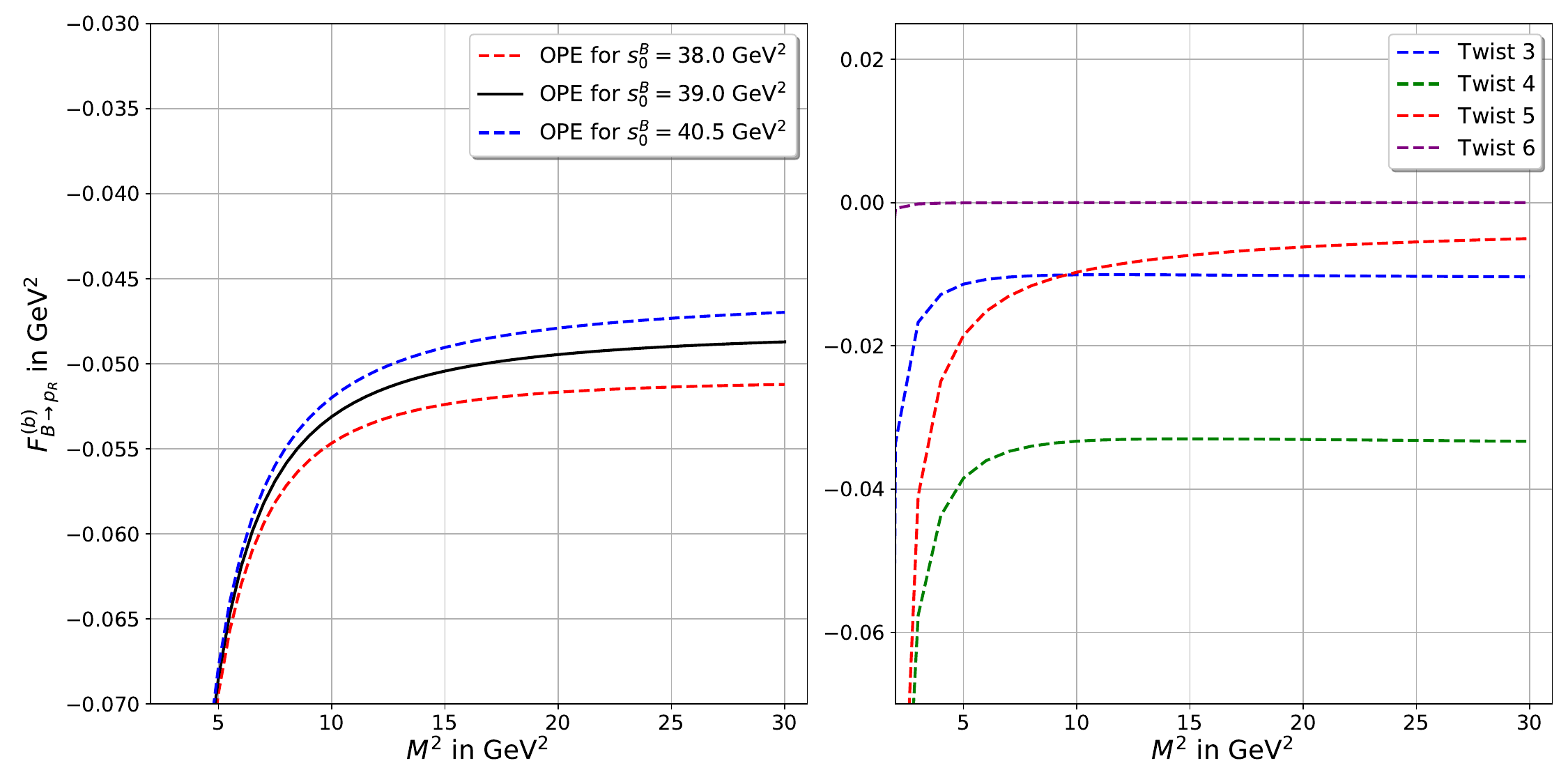}
	}
	\caption{Analysis for the form factor $F_{B \to p_R}^{(b)}(q^2)$ in the $(b)$-model for different Borel parameters $M^2$. The left plot shows all twist contributions combined for various choices of the threshold parameter $s_0^B$, while the right plot illustrates each twist contribution individually. The mass of the dark matter particle $m_{\Psi}$ is set to the benchmark value $m_{\Psi} = 2$ GeV \cite{Alonso-Alvarez:2021qfd}.}
	\label{fig:FBpRb}
\end{figure}

\begin{figure}[H]
	\centerline{
		\includegraphics[width=14cm, height=7cm]{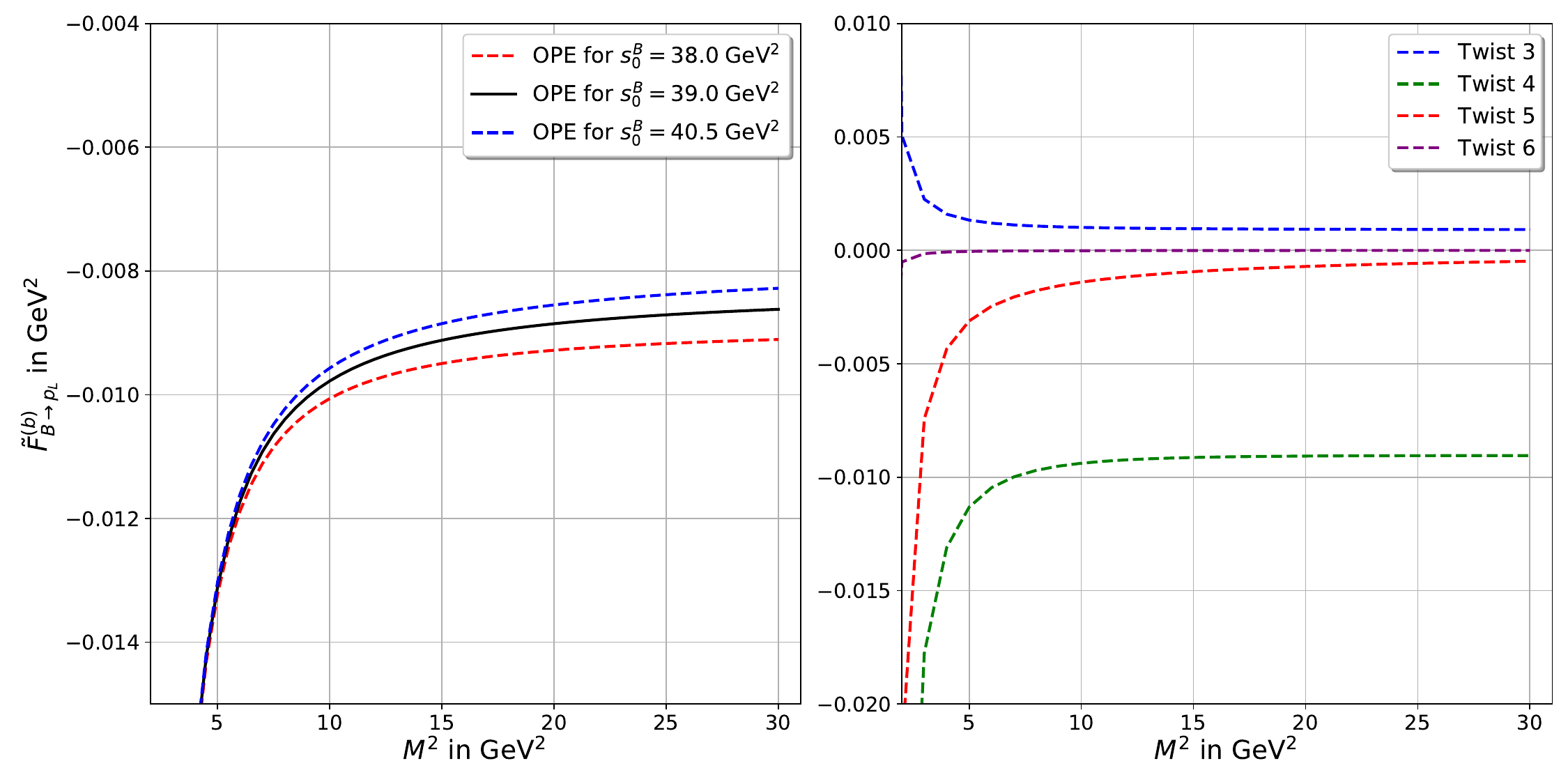}
	}
	\caption{Similar analysis as in figure \ref{fig:FBpRd} for the form factor $\widetilde{F}_{B \to p_L}^{(b)}(q^2)$.}
	\label{fig:FBpLb}
\end{figure}

Next we determine the uncertainties around the central values of the form factors. This requires to consider the form factors before their z-expansion and we individually introduce variations to each input parameter within its predefined range of uncertainty. Subsequently, we perform a $z$-expansion for each parameter variation yielding two sets of parameters for the slope parameter $b_{B\to p_R}^{(d)}$ and the normalization $F_{B\to p_R}^{(d)}(0)$, namely the upper and lower bound on these parameters with respect to the input parameter variation. The values in each set are then added in quadrature to obtain the possible parameter range for both fitting parameters. Apart from the usual correlation between the threshold parameter $s_0^B$ and the Borel parameter $M^2$ and the impact of the $\mu$ variation on the renormalization scale dependent parameters, we assume that the remaining parameters are completely uncorrelated.
\noindent
We state the slope parameters and the form factor at $q^2 = 0$ within their uncertainties in table \ref{tab:FitParametersModeld} for the $(d)$-model and in table \ref{tab:FitParametersModelb} for the $(b)$-model: \\
\begin{table}[H]
	\centering
	\begin{tabular}{|c|c||c|c|}
		\hline
		&&&\\[-2mm]
		$F^{(d)}_{B\to p_R}(0)$& $b^{(d)}_{B\to p_R} $ & $\widetilde{F}^{(d)}_{B\to p_L}(0)$& $b^{(d)}_{B\to p_L}$ \\
		&&&\\[-2mm]
		\hline
		\hline
		&&&\\[-2mm]
		$0.022^{+0.013}_{-0.013} $& $ 4.46^{+0.97}_{-1.72} $& $ 0.005_{-0.001}^{+0.002} $&$ -2.27_{-0.08}^{+0.10}$ \\[2mm]
		\hline
	\end{tabular}
	\caption{ Parameters of the $z$-expansion in the case the $(d)$-model form factors (in GeV$^2$ units) including all contributions to the nucleon DA up to twist six.} 
	\label{tab:FitParametersModeld}
\end{table}
\begin{table}[H]
	\centering
	\begin{tabular}{|c|c||c|c|}
		\hline
		&&&\\[-2mm]
		$F^{(b)}_{B\to p_R}(0)$& $b^{(b)}_{B\to p_R} $ & $\widetilde{F}^{(b)}_{B\to p_L}(0)$& $b^{(b)}_{B\to p_L}$ \\
		&&&\\[-2mm]
		\hline
		\hline
		&&&\\[-2mm]
		$ -0.041_{-0.018}^{+0.019} $ & $ -2.00_{-3.62}^{+1.58}$ & $-0.007_{-0.002}^{+0.003}$ & $ -2.85_{-0.15}^{+0.17}$ \\[2mm]
		\hline
	\end{tabular}
	\caption{ Parameters of the $z$-expansion for the $(b)$-model form factors (in GeV$^2$ units) including all contributions to the nucleon DA up to twist six.} 
	\label{tab:FitParametersModelb}
\end{table}\noindent
Now one can use these input parameters to extrapolate the four different form factors and obtain these for the two models. \\
Figures \ref{fig:FBpRd} to \ref{fig:FBpLb} show the form factors with respect to the Borel parameter $M^2$. The right panel depicts the individual twist contributions to the different form factors such that a direct comparison between the higher twist corrections and the leading twist-three contribution becomes possible. These leading contributions have been previously examined in \cite{Khodjamirian:2022vta} and can also be found in \cite{Elor:2022jxy}. In this context, it is notable that the leading contribution to the form factor \(\widetilde{F}^{(d)}_{B \rightarrow p_{L}}(q^2)\) starts at twist four accuracy, contrary to the other three form factors. Nevertheless, the OPE shows in general good convergence for all form factors, which is in accordance with other sum rule analyses including $B$-meson interpolating currents. \\
 \noindent
 \noindent
Although the OPE shows good convergence for all form factors, we observe at the benchmark value $m_\Psi = 2$ GeV that the $(b)$-model exhibits a substantial twist-four correction, violating the typical hierarchy of the OPE since the twist-four contribution is larger compared to the leading-twist contribution. This has its origin in the significant $\mathcal{T}_{2,4}$ contributions in the $(b)$-model, which vanish in the $(d)$-model computation. Nevertheless, the convergence of the OPE is still well established beyond twist four \sout{in the $(b)$-model}. \\
In the left panel, we investigate the complete form factor expressions including all contributions up to twist six at the benchmark value $m_\Psi = 2$ GeV. At this stage, it is interesting to compare these expressions for different choices of the threshold parameter $s_0^B$ based on the upper and lower bounds specified in table  \ref{tab:input}. Given the small fluctuations of the form factors with respect to different Borel parameters across varying thresholds, we see that the sum rules for the four form factors remain stable at the reference value $m_\Psi=2 $ GeV, thus ensuring their reliability. \\
 \noindent
The choice of the Borel parameter range $M^2$ aligns with the specifications from table \ref{tab:input} and is in agreement with the findings in \cite{Khodjamirian:2022vta}. Furthermore, an alternative validation of this Borel window can be performed by considering the form factors over a broader range of the Borel parameter $M^2$ and determining the Borel window based on the stability of the sum rule with respect to $M^2$. As illustrated in figures \ref{fig:FBpRd} to \ref{fig:FBpLb}, it is evident that the sum rules remain stable within the specific Borel window defined in table \ref{tab:input}. Additionally, this choice of the Borel parameter window ensures that continuum and excited states are reasonably suppressed and contribute approximately around $20\%$ to $30\%$. Consequently, our findings are not excessively influenced by the Quark-Hadron Duality (QHD) approximation. \\
Moreover, the determination of the threshold parameter $s_0^B$ is accomplished by calculating the derivative of the sum rules presented in Eq. \eqref{eq:LCSR} with respect to $-1/M^2$. By taking the ratio between the derivative outcome with Eq. \eqref{eq:LCSR}, we obtain an approximation for the $B$-meson mass $m_B$, which is subsequently adjusted by choosing the value of $s_0^B$ such that it fits existing literature data \cite{ParticleDataGroup:2018ovx}. The stability of the sum rules in figures \ref{fig:FBpRd} to \ref{fig:FBpLb} shows again the validity of this approach. \\
\begin{figure}[H]
	\centerline{
		\includegraphics[width=14cm, height=7cm]{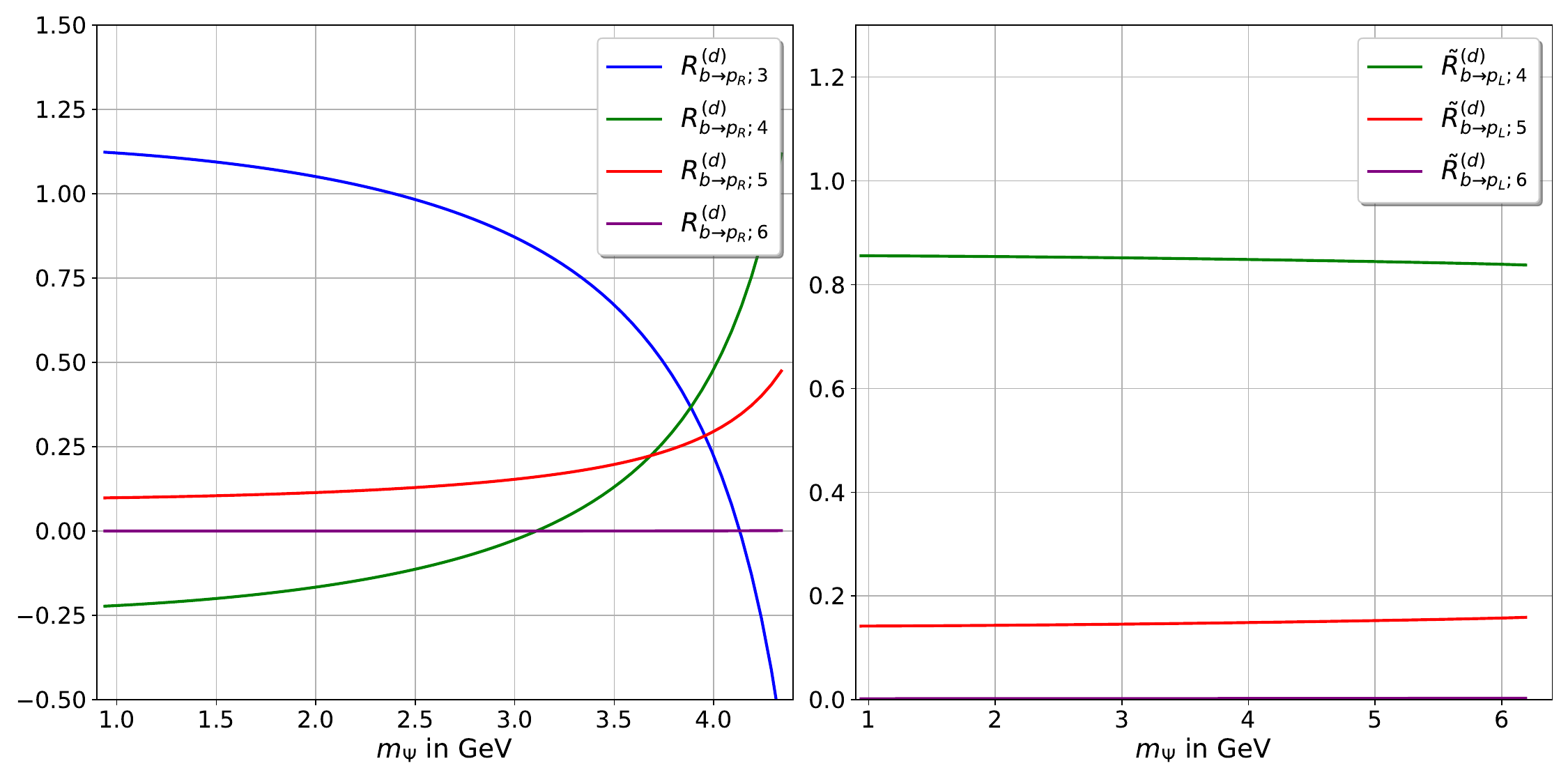}
	}
	\caption{Plot for the ratios $R_{B \to p_R;i}^{(d)}$ and $\widetilde{R}_{B \to p_L;i}^{(d)}$ defined in Eq. \eqref{eq:RatioFBpRd} and \eqref{eq:RatioFBpRb} with respect to the dark fermion mass $m_{\Psi}$. The right panel contains the ratios for the form factor $\widetilde{F}^{(d)}_{B \to p_L}$, while The left panel shows the contributions from the form factor $F^{(d)}_{B \to p_R}$.}
	\label{fig:mPsiRatioFBd}
\end{figure} 
\noindent
So far, our discussion has been centered around the chosen value $m_{\Psi} = 2$ GeV. Since the LCSR approach works in the limit $q^2 = m_{\Psi}^2 \ll m_b^2$, we naturally assume that the OPE also converges when $m_{\Psi}$ remains below 2 GeV. It is important to understand the applicability of the OPE across higher values of $m_{\Psi}$, especially at which point the OPE starts to break down. This analysis is shown for the $(d)$-model in figure \ref{fig:mPsiRatioFBd}, focusing on the form factor $F_{B \to p_R}^{(d)}(q^2)$ in the left panel and on the form factor $\widetilde{F}_{B \to p_L}^{(d)}(q^2)$ in the right panel. For these investigation, it is useful to introduce the ratios:
\begin{figure}[t]
	\centerline{
		\includegraphics[width=14cm, height=7cm]{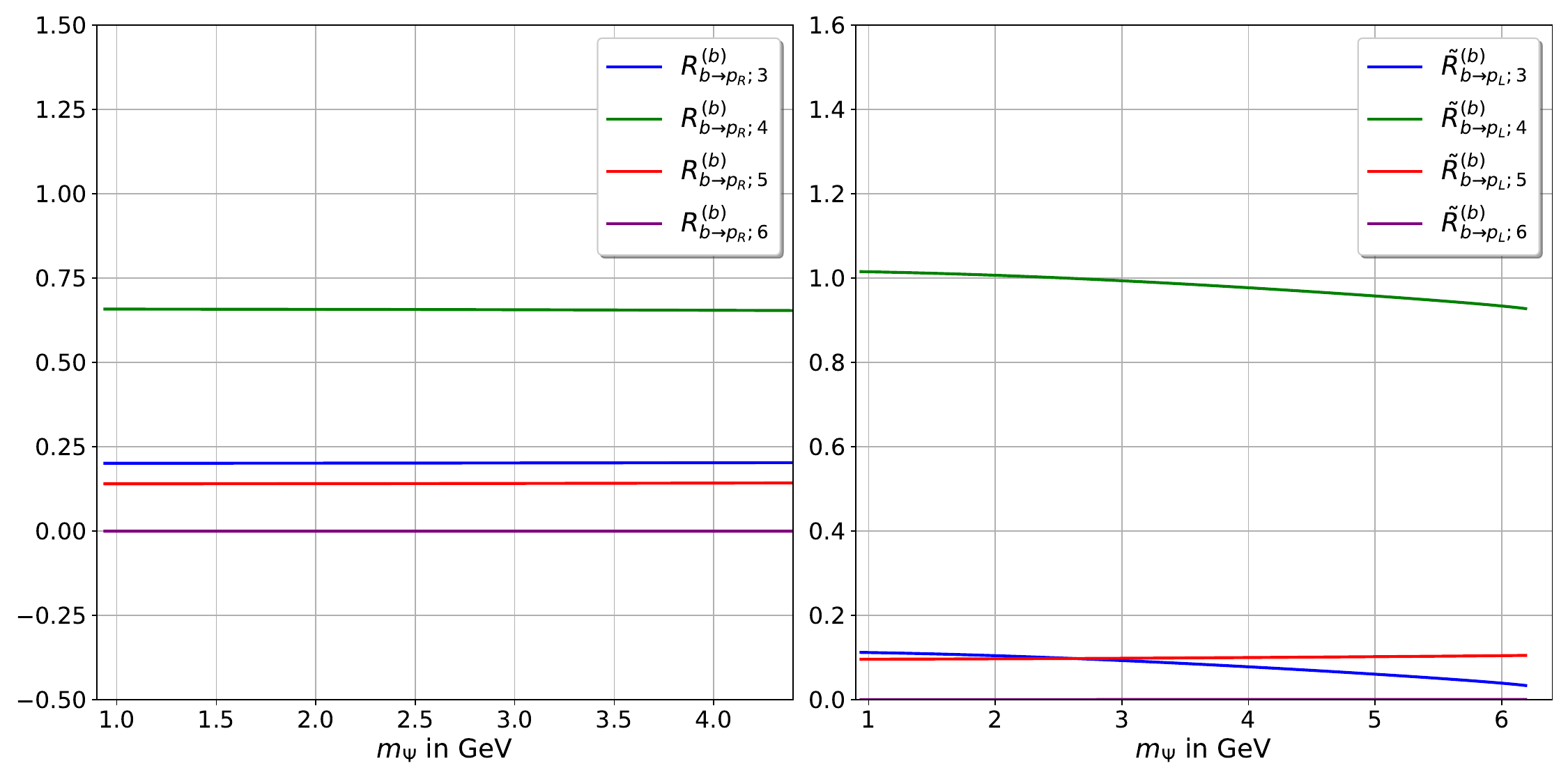}
	}
	\caption{Plot for the ratios $R_{B \to p_R;i}^{(b)}$ and $\widetilde{R}_{B \to p_L;i}^{(b)}$ defined in Eq. \eqref{eq:RatioFBpRd} and \eqref{eq:RatioFBpRb} after replacing $(d) \to (b)$ with respect to the dark fermion mass $m_{\Psi}$. The right panel contains the ratios for the form factor $\widetilde{F}^{(b)}_{B \to p_L}$, while The left panel shows the contributions from the form factor $F^{(b)}_{B \to p_R}$.}
	\label{fig:mPsiRatioFBb}
\end{figure} \noindent
\begin{align}
	 R_{B \to p_R;i}^{(d)} = \, \frac{F^{(d)}_{B \to p_R;i}}{\sum_{i \in \{3,4,5,6\}}F^{(d)}_{B \to p_R;i}} \, ;  \widetilde{R}_{B \to p_L;i}^{(d)} = \, \frac{\widetilde{F}^{(d)}_{B \to p_L;i}}{\sum_{i \in \{4,5,6\}} \widetilde{F}^{(d)}_{B \to p_L;i}} , 	\label{eq:RatioFBpRd} \\
	R_{B \to p_R;i}^{(b)} = \, \frac{F^{(b)}_{B \to p_R;i}}{\sum_{i \in \{3,4,5,6\}}F^{(b)}_{B \to p_R;i}} \, ;  \widetilde{R}_{B \to p_L;i}^{(b)} = \, \frac{\widetilde{F}^{(b)}_{B \to p_L;i}}{\sum_{i \in \{3,4,5,6\}} \widetilde{F}^{(b)}_{B \to p_L;i}} , 	\label{eq:RatioFBpRb}
\end{align}
where $F^{(d)}_{B \to p_R;i}$ and $\widetilde{F}^{(d)}_{B \to p_L;i}$ belong to the twist $i$ contribution of the form factors $F^{(d)}_{B \to p_R}$ and $\widetilde{F}^{(d)}_{B \to p_L}$.\\ 
Using \eqref{eq:RatioFBpRd} and \eqref{eq:RatioFBpRb}, we derive an approximation for $m_\Psi$ where the OPE diverges. The underlying notion here is that when the higher twist contributions grow, which is indicated by increased ratios $R_{B \to p_R;i}^{(d)}$ and $\widetilde{R}_{B \to p_L;i}^{(d)}$, these contributions become dominant and spoil the convergence of the OPE in general. Our observations reveal that for the form factor $F_{B \to p_R}^{(d)}(q^2)$ in the left panel of figure \ref{fig:mPsiRatioFBd}, the convergence is spoiled around $m_\Psi \approx 3$ GeV. In comparison to that the form factor $\widetilde{F}^{(d)}_{B \to p_L}(q^2)$ turns out to be insensitive to this investigation. Nonetheless, for the earlier considered benchmark value of 2 GeV, the convergence of the OPE remains robust. It is important to note that probing $m_\Psi$ beyond 6.2 GeV is unfeasible since we cross the multihadron threshold at $t_+$, which results into complex $z$-parameter values. \\
A similar examination can be carried out for the $(b)$-operator and this is shown in figure \ref{fig:mPsiRatioFBb}. This underlines that the hierarchy in the twist expansion remains intact for both form factors confirming that the expansion retains its convergence throughout the entire kinematical range of $m_\Psi$.

\subsection{Branching fractions} \label{chp:BranchingFractions}

\begin{figure}[H]
	\centerline{
		\includegraphics[width=10cm, height=7.5cm]{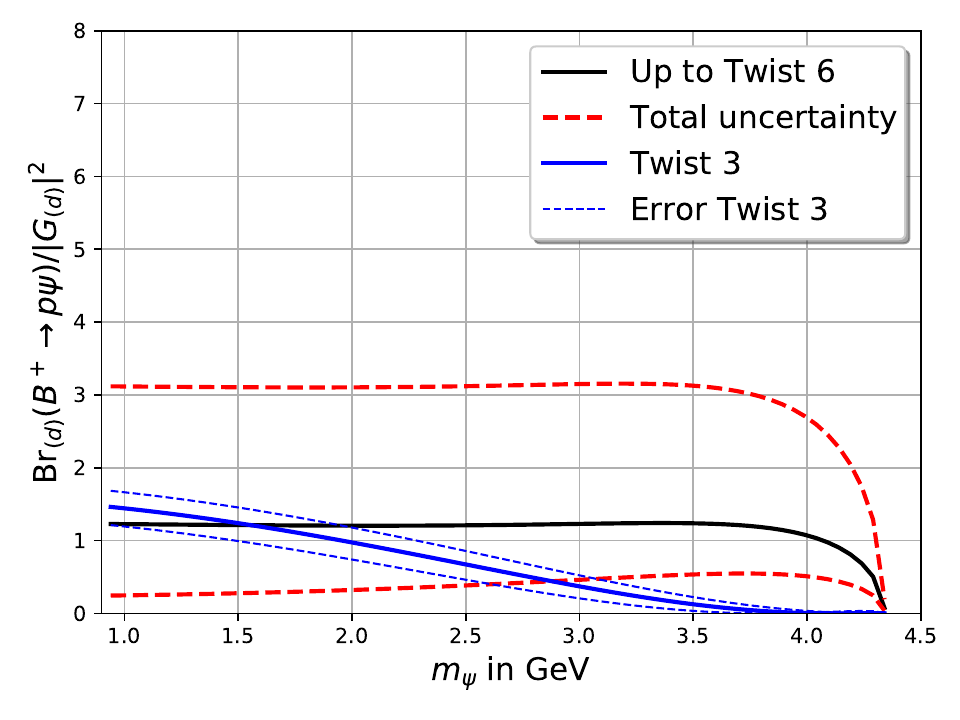}
	}
	\caption{Branching fraction for the decay $B \to p \Psi$ in the $(d)$-model with respect to the dark matter particle mass $m_{\Psi}$, normalized to the effective four-fermion coupling $|G_{(d)}|^2$. The blue line with the blue dashed error band illustrates the original twist-three computation from \cite{Khodjamirian:2022vta}, while the black curve shows the computation including contributions up to twist six. The dashed red curves represents the uncertainty on this calculation.}
	\label{fig:BRd}
\end{figure}

Before calculating the branching fraction, we need to obtain the decay amplitude of the considered $B^+\to p\Psi$ decay. 
Following \cite{Khodjamirian:2022vta}, we start with the amplitude ${\cal A}_{(d)}(B^+\to p \Psi)$ from Eq. \eqref{eq:AmplitudeStart} and insert the decomposition into form factors for the $B \to p$ transition from Eq. \eqref{eq:ffDecomp}
\begin{align}
	\mathcal{A}_{(d)}(B^+ \to p \Psi) = \, G_{(d)} \bar{u}_{p,R}(P) \Big[A^{(d)} + B^{(d)} \gamma_5\Big] u_{\Psi}^c(q) \; ,\label{eq:MatrixElementFinal}
\end{align} 
with
\begin{align}
	A^{(d)} \equiv& \; \frac12 \Bigg[F^{(d)}_{B\to p_R}(q^2) + \frac{m_{\Psi}}{m_p} \widetilde{F}^{(d)}_{B\to p_R}(q^2)\Bigg] + \frac12 \Bigg[F^{(d)}_{B\to p_L}(q^2) + \frac{m_{\Psi}}{m_p} \widetilde{F}^{(d)}_{B\to p_L}(q^2)\Bigg] \\
	B^{(d)} \equiv& \; -\frac12 \Bigg[F^{(d)}_{B\to p_R}(q^2) + \frac{m_{\Psi}}{m_p} \widetilde{F}^{(d)}_{B\to p_R}(q^2)\Bigg] + \frac12 \Bigg[F^{(d)}_{B\to p_L}(q^2) + \frac{m_{\Psi}}{m_p} \widetilde{F}^{(d)}_{B\to p_L}(q^2)\Bigg] \, ,
\end{align}
We have expressed the decay amplitude through the four different form factors and the ratios of the proton mass $m_p$ as well as the dark matter particle  mass $m_\Psi$.
In order to obtain the two-body decay width, one has to square the amplitude from Eq. \eqref{eq:MatrixElementFinal} and multiply this expression with normalisation factors
\begin{align}
	\Gamma_{(d)}(B^+ \to p \Psi) = \frac{1}{2 m_B} \int \Md \Pi \, |\mathcal{A}_{(d)}(B^+ \to p \Psi)|^2 \, .
\end{align}
The phase space integration for a two-body decay can be carried out analytically yielding the K\"allen function
\begin{align}
	\Gamma_{(d)}(B^+ \to p \Psi) 
	=& \; |G_{(d)}|^2  \bigg[|A^{(d)}|^2 (m_B^2 - (m_p - m_{\Psi})^2) + |B^{(d)}|^2 (m_B^2 - (m_p + m_{\Psi})^2)\bigg] \cdot \nonumber \\& \frac{\lambda^{\frac12}(m_B^2, m_p^2, m_{\Psi}^2)}{8 \pi m_B^3} \; . \label{eq:DecayWidthModeld}
\end{align}
The final expression for the decay width in terms of form factors is given by
\begin{align}
	&\Gamma_{(d)}(B^+\to p\Psi) =
	|G_{(d)}|^2 \Bigg\{\Bigg[\Big(F^{(d)}_{B\to p_R}(m_\Psi^2)\Big)^2
	+ \frac{m_\psi^2}{m_p^2}\Big(\widetilde{F}^{(d)}_{B\to p_L}(m_\Psi^2)\Big)^2\Bigg] \nonumber \\
	&\times \big(m_B^2-m_p^2-m_\Psi^2\big)
	+ \;
	2 m_\Psi^2F^{(d)}_{B\to p_R}(m_\Psi^2)\widetilde{F}^{(d)}_{B\to p_L}(m_\Psi^2)\Bigg\}
	\,\frac{\lambda^{1/2}(m_B^2,m_p^2,m_\Psi^2)}{16\pi m_B^3}\,.
	\label{eq:widthd}
\end{align}\noindent
In comparison to the corresponding expression in \cite{Khodjamirian:2022vta,Elor:2022jxy}, the form factor $\widetilde{F}^{(d)}_{B\to p_L}(q^2)$ contributes here due to higher twist contributions. Since this form factor already appears in the $(b)$-model at the leading-twist approximation, a simple exchange of $(d)$ with $(b)$ allows us to derive corresponding relations for the second model. \\
The observable of interest is the branching fraction for this decay. This parameter can be derived by dividing Eq. \eqref{eq:widthd} by the total width of the $B$ meson. Alternatively, we can also multiply Eq. \eqref{eq:widthd} by the $B$-meson lifetime $\tau_{B^\pm}=1.638 \pm 0.004$ ps from \cite{ParticleDataGroup:2020ssz}:
\begin{align}
	\mathrm{Br}_{(d)}(B^+\to p\Psi) = \Gamma_{(d)}(B^+\to p\Psi) \cdot \tau_{B^\pm} \;.
	\label{eq:Branchingd}
\end{align}\noindent
As shown in figure \ref{fig:BRd}, we present the branching fraction for the $(d)$-model incorporating contributions to the nucleon distribution amplitudes up to twist six and compare them to the leading-twist contribution from \cite{Khodjamirian:2022vta,Elor:2022jxy}. Within $m_{\Psi}$ values up to 3 GeV, we identify that both computations agree very well within their uncertainties. This observation aligns with our earlier observation in figure \ref{fig:mPsiRatioFBd} that at $m_{\Psi} = 3$ GeV the higher twist corrections become dominant and ultimately affect the convergence of the OPE.\\
Generally, the uncertainties on our twist six calculation turn out to be larger compared to the previous leading-twist evaluation. This disparity arises due to the larger error estimates on input parameters of the distribution amplitudes in the conformal and next-to-conformal expansion. In particular, the parameter $\xi_{10}$ introduces large uncertainties as we assume a conservative error of $50\%$ based on the value from \cite{Braun:2006hz}. Nonetheless, we affirm in general that the leading-twist outcome constitutes a reliable approximation for the branching fractions within the $m_\Psi$-range up to 3 GeV. \\
However, a significant discrepancy between the twist-three calculation from \cite{Khodjamirian:2022vta,Elor:2022jxy} and our computation arises for the $(b)$-model. In this case, the branching fraction increases by roughly a factor of 20 and the two computations do not agree within their uncertainties. This deviation can be attributed to the substantial $\mathcal{T}_{2,4}$-contributions at the twist four level which we observe for both form factors, as it is evident from figures \ref{fig:FBpRb} and \ref{fig:FBpLb}. \\
Similar to the $(d)$-model, the uncertainties on the twist six computation are notable and share the same origin as for the $(d)$-model. Particularly, the upper bound uncertainty becomes enhanced once $m_{\Psi} > 3$ GeV. This behavior illustrates that the branching fraction loses its reliability as $m_{\Psi}^2 \sim m_b^2$, thereby violating a crucial requirement for the light-cone expansion. 
 \noindent
Consequently, we conclude that the leading-twist analysis from \cite{Khodjamirian:2022vta} for this specific model falls short as higher twist corrections shift the value of the branching fractions considerably and spoil the hierarchy of the OPE. However, the branching fraction now lies with these additional contributions in the sensitivity range of Belle-II, estimated to be around $3 \cdot 10^{-6}$ \cite{Alonso-Alvarez:2021qfd}.
\begin{figure}[t]
	\centerline{
		\includegraphics[width=10cm, height=7.5cm]{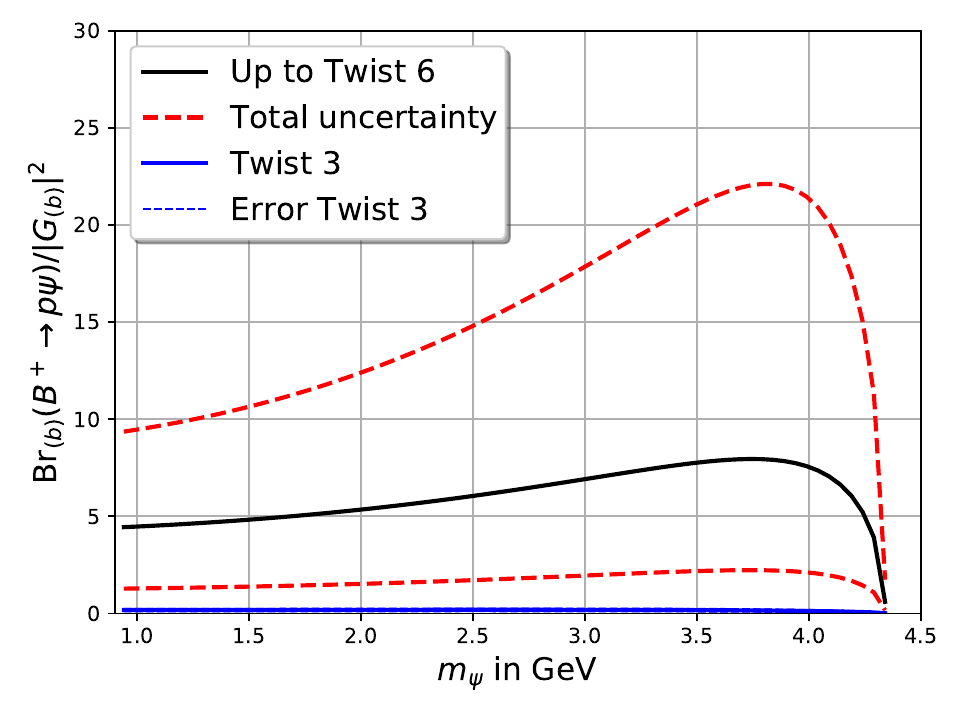}
	}
	\caption{We show the same as in figure \ref{fig:BRd}, but for the $(b)$-model.}
	\label{fig:BRb}
\end{figure} \noindent

\section{Conclusion}  \label{chp:Conclusion}
The SM plays a crucial role in modern particle physics, although there are effects like dark matter or the baryon asymmetry of the universe which are not incorporated into this theory. As there are strong experimental hints for the existence of dark matter, several models are proposed of which we have studied the recently proposed $B$-Mesogenesis model \cite{Alonso-Alvarez:2021oaj,Alonso-Alvarez:2021qfd,Elor:2018twp}. It is especially interesting for experimental facilities like Belle-II since it predicts new dark matter particles at energy scales which are in principle within its sensitivity range. \\
We have focused here on one allowed decay mode within this model, namely the decay $B \to p \Psi$. 
For this, we have used the light-cone sum rule approach to determine the branching fraction of the decay $B \to p \Psi$. While the leading-twist contributions have been obtained in previous works \cite{Khodjamirian:2022vta, Elor:2022jxy}, we include higher twist corrections up to twist six into our analysis in order to check the reliability of the leading-twist analysis and the convergence of the operator product expansion in general. Thus, we have followed the procedure from \cite{Khodjamirian:2022vta} and computed the branching fractions of the two considered versions (the $(d)$- and the $(b)$-model) in the $B$-Mesogenesis scenario, normalized to the effective four-fermion couplings $G_{(d)}$ and $G_{(b)}$. \\ \noindent
We observe for the two form factors of the $(d)$-model, which contribute to the respective branching fraction, that higher twist contributions become increasingly smaller in the parameter range $0.94$ GeV $\le m_{\Psi} \le$ 3 GeV, indicating that the OPE converges and that the leading-twist contribution is dominant. This shows that the higher twist corrections have a minimal impact on the branching ratio in this parameter range and that the behavior mostly follows \cite{Khodjamirian:2022vta}. \\
\noindent
Beyond 3 GeV, we see that the that higher twist corrections start to dominate and therefore that the OPE breaks down. This behavior is expected since the light-cone approach requires that $m_{\Psi}^2 \ll m_b^2$, which starts to get violated beyond 3 GeV. Hence, we conclude that the branching fraction estimate from \cite{Khodjamirian:2022vta} is reliable in the regime $m_{\Psi} \le 3$ GeV and in agreement with the results from this work. \\
\noindent
In contrast to that the twist four contributions tend to be the dominant contribution in the $(b)$-model, albeit the OPE converges beyond this twist correction. This has the consequence that the branching fraction for this model is significantly increased, namely by factor of $20$. This observation has a direct impact on experimental searches, since both versions of the $B$-Mesogenesis model are now in the sensitivity range of Belle-II.\\
\noindent
From an experimental point of view, the decays $B\to \Lambda \Psi \to p \pi \Psi$ or $B \to \Lambda \Psi \to p \pi \Psi$ are however easier accessible as they see two SM particles in the final state combined with the new-physics particle $\Psi$ in form of missing energy in the detector. But the necessary distribution amplitudes for the $\Delta$- and $\Lambda$-baryons are less known and only a leading-twist analysis is available in the literature. Nevertheless, ratios of these decays with the computed branching fraction $B \to p \Psi$ in this work would be independent of the couplings $G_{(d)}$ and $G_{(b)}$ and hence reduce the number of input parameters.

\section*{Acknowledgements}
We thank A. Khodjamirian for useful discussions and T. Mannel for reading the manuscript. This work was supported by the Deutsche Forschungsgemeinschaft (DFG, German Research Foundation) under the grant 396021762 - TRR 257.

\newpage

\appendix

\section{Nucleon distribution amplitude} \label{chp:NucleonDA}

Generally, the hadronic matrix element from the previous discussion can be decomposed into different Lorentz structures based on symmetry considerations based on Lorentz covariance, spin and parity of the proton \cite{Braun:2000kw}:
\begin{eqnarray}
	\label{eq:DecomNuclDAs}
	\lefteqn{ 4 \bra{0} \epsilon^{ijk}
		u_\alpha^i(a_1 x) u_\beta^j(a_2 x) d_\gamma^k(a_3 x) \ket{p(P)} }
	\nonumber \\
	&=& {\cal S}_1 m_p C_{\alpha \beta} \left(\gamma_5 u_p(P)
	\right)_\gamma + {\cal S}_2 m_p^2 C_{\alpha \beta}
	\left(\!\not\!{x} \gamma_5 u_p(P) \right)_\gamma + {\cal P}_1 m_p
	\left(\gamma_5 C\right)_{\alpha \beta} {(u_p(P))}_\gamma \nonumber \\
	&& + {\cal P}_2 m_p^2 \left(\gamma_5 C \right)_{\alpha \beta}
	\left(\!\not\!{x} u_p(P) \right)_\gamma
	+ \left(\mathcal{V}_1+\frac{x^2m_p^2}{4}\mathcal{V}_1^M \right)
	\left(\!\not\!{P}C \right)_{\alpha \beta} \left(\gamma_5 u_p(P)
	\right)_\gamma \nonumber \\
	&& + {\cal V}_2 m_p \left(\!\not\!{P} C \right)_{\alpha \beta}
	\left(\!\not\!{x} \gamma_5 u_p(P) \right)_\gamma + {\cal V}_3 m_p
	\left(\gamma_\mu C \right)_{\alpha \beta}\left(\gamma^{\mu}
	\gamma_5 u_p(P) \right)_\gamma
	\nonumber \\
	&& + {\cal V}_4 m_p^2 \left(\!\not\!{x}C \right)_{\alpha \beta}
	\left(\gamma_5 u_p(P) \right)_\gamma + {\cal V}_5 m_p^2
	\left(\gamma_\mu C \right)_{\alpha \beta} \left(i \sigma^{\mu\nu}
	x_\nu \gamma_5 u_p(P) \right)_\gamma \nonumber \\
	&& + {\cal V}_6 m_p^3 \left(\!\not\!{x} C \right)_{\alpha \beta}
	\left(\!\not\!{x} \gamma_5 u_p(P) \right)_\gamma  +
	\left(\mathcal{A}_1+\frac{x^2m_p^2}{4}\mathcal{A}_1^M\right)
	\left(\!\not\!{P}\gamma_5 C \right)_{\alpha \beta} {(u_p(P))}_\gamma
	\nonumber \\
	&& + {\cal A}_2 m_p \left(\!\not\!{P}\gamma_5 C \right)_{\alpha
		\beta} \left(\!\not\!{x} u_p(P) \right)_\gamma  + {\cal A}_3 m_p
	\left(\gamma_\mu \gamma_5 C \right)_{\alpha \beta}\left(
	\gamma^{\mu} u_p(P) \right)_\gamma
	\nonumber \\
	&& + {\cal A}_4 m_p^2 \left(\!\not\!{x} \gamma_5 C \right)_{\alpha
		\beta} {(u_p(P))}_\gamma + {\cal A}_5 m_p^2 \left(\gamma_\mu \gamma_5
	C \right)_{\alpha \beta} \left(i \sigma^{\mu\nu} x_\nu u_p(P)
	\right)_\gamma \nonumber \\
	&& + {\cal A}_6 m_p^3 \left(\!\not\!{x} \gamma_5 C \right)_{\alpha
		\beta} \left(\!\not\!{x} u_p(P) \right)_\gamma  +
	\left(\mathcal{T}_1+\frac{x^2m_p^2}{4}\mathcal{T}_1^M\right)
	\left(P^\nu i \sigma_{\mu\nu} C\right)_{\alpha \beta} \nonumber \\
	&& \times
	\left(\gamma^\mu\gamma_5 u_p(P) \right)_\gamma  + {\cal T}_2 m_p \left(x^\mu P^\nu i \sigma_{\mu\nu}
	C\right)_{\alpha \beta} \left(\gamma_5 u_p(P) \right)_\gamma + {\cal
		T}_3 m_p \left(\sigma_{\mu\nu} C\right)_{\alpha \beta} \nonumber \\ && \times
	\left(\sigma^{\mu\nu}\gamma_5 u_p(P) \right)_\gamma  + {\cal T}_4 m_p \left(P^\nu \sigma_{\mu\nu} C\right)_{\alpha
		\beta} \left(\sigma^{\mu\rho} x_\rho \gamma_5 u_p(P) \right)_\gamma \nonumber \\ && +  
	{\cal T}_5 m_p^2 \left(x^\nu i \sigma_{\mu\nu} C\right)_{\alpha
		\beta} \left(\gamma^\mu\gamma_5 u_p(P) \right)_\gamma  + {\cal T}_6 m_p^2 \left(x^\mu P^\nu i \sigma_{\mu\nu}
	C\right)_{\alpha \beta} \left(\!\not\!{x} \gamma_5 u_p(P)
	\right)_\gamma \nonumber \\
	&& + {\cal T}_{7} m_p^2 \left(\sigma_{\mu\nu}
	C\right)_{\alpha \beta}  \left(\sigma^{\mu\nu} \!\not\!{x} \gamma_5
	u_p(P) \right)_\gamma  + {\cal T}_{8} m_p^3 \left(x^\nu \sigma_{\mu\nu}
	C\right)_{\alpha \beta} \left(\sigma^{\mu\rho} x_\rho \gamma_5 u_p(P)
	\right)_\gamma \nonumber  \\ 
\end{eqnarray} \noindent
On the left side of Eq. \eqref{eq:DecomNuclDAs}, we investigate a proton to vacuum matrix element with an on-shell proton of $P^2 = m_p^2$ in the initial state. The quark fields $u(x), u(0)$ and $d(0)$ correspond to the valence quarks inside the proton. The Greek letters $\alpha, \beta, \gamma$ denote spinor indices, while Latin letters  $i,j,k$ are colour indices. 
Gauge link factors between each valence quark are suppressed rendering the expression in Eq. \eqref{eq:DecomNuclDAs} gauge invariant. 
According to the previous discussion in \cite{Khodjamirian:2022vta}, we can set $a_1 = 1$ and $a_2 = a_3 = 0$ in our case and note that the matrix $C$ is the charge conjugation matrix defined as $C = \gamma_2 \gamma_0$ and $u_p(P)$ is the proton spinor. The tensor $\sigma_{\mu \nu}$ is defined in terms of $\sigma_{\mu \nu} = \frac{i}{2} [\gamma_{\mu}, \gamma_{\nu}]$. \\
As Eq. \eqref{eq:DecomNuclDAs} indicates, there are in total 24 invariant functions $\mathcal{S}_i$, $\mathcal{P}_i$, $\mathcal{A}_i$, $\mathcal{V}_i$, $\mathcal{T}_i$, which we can not assign a definite twist. 
\noindent
These calligraphic quantities can be related to the twist amplitudes in the following way:

\begin{equation}
	\mathcal{F}(a_1,a_2,a_3,(P\cdot x))= \int \! \Md \alpha_1 \Md \alpha_2 \Md \alpha_3 \;
	\delta(1\!-\!\alpha_1\!-\!\alpha_2\!-\!\alpha_3) e^{-i
		(P  \cdot  x) \sum_i \alpha_i a_i} F(\alpha_i)
	\label{eq:CalRelation}
\end{equation} \noindent
The variables $\alpha_{1,2,3}$ denote the momentum fractions of the different quarks inside the proton. We start with twist three contributions and relate the calligraphic quantities appearing in Eq. \eqref{eq:DecomNuclDAs} with Eq. \eqref{eq:CalRelation} to the definite twist amplitudes:
\begin{table}[H]
	\begin{center}
		\begin{tabular}{|c|c|}
			\hline
			&\\[-4.0mm]
			$\mathcal{F}$  & integrand on r.h.s. of (\ref{eq:CalRelation}) \\
			&\\[-4.0mm]
			\hline
			\hline 
			&\\[-4.0mm]
			${\cal V}_1$ & $V_1$ \\
			&\\[-4.0mm]
			\hline 
			&\\[-4.0mm]
			${\cal A}_1$ & $A_1$ \\
			&\\[-4.0mm]
			\hline 
			&\\[-4.0mm]
			${\cal T}_1$ & $T_1$ \\[1mm]
			\hline
		\end{tabular}
	\end{center}
\end{table} \noindent
In the above table, the definite twist distribution amplitudes are given by
\begin{eqnarray}
	V_1(\alpha_i) =& 120 \alpha_1 \alpha_2 \alpha_3 [  \phi_3^0  +
	\phi_3^+ (1- 3 \alpha_3) ] \,, \nonumber \\
	A_1(\alpha_i) =& 120 \alpha_1 \alpha_2 \alpha_3 (\alpha_2 - \alpha_1)\phi_3^- \,, \nonumber \\
	T_1(\alpha_i) =& 120 \alpha_1 \alpha_2 \alpha_3 [  \phi_3^0  - {1 \over
		2}(\phi_3^+-\phi_3^-) (1- 3 \alpha_3) ] \,. \label{eq:tw3DA}
\end{eqnarray}
At leading twist, there are in total three coefficients $\phi_3^{(0,\pm)}$, which can be parameterized through the parameters $f_N$, the normalization factor in leading conformal spin, and $A_1^u$ as well as $V_1^d$, which belong both to the next to leading conformal spin:
\begin{align}
	\phi_3^0 = f_N; \; \; \; \phi_3^+ = \frac72 f_N (1 - 3V_1^d); \; \; \; \phi_3^- = \frac{21}{2} f_N A_1^u
\end{align}
The remaining contributions can be classified according to their twist and specify their tensor structure \cite{Braun:2000kw}:
\begin{table}[H]
	\begin{center}
		\begin{tabular}{|c||c|c|c|c|}
			\hline
			&&&&\\[-4.0mm]
			& twist 3 & twist 4 & twist 5 & twist 6 \\
			&&&&\\[-4.0mm]
			\hline
			\hline
			&&&&\\[-4.0mm]
			vector & $V_1$ & $V_2, V_3$ & $V_4, V_5$ & $V_6$ \\
			&&&&\\[-4.0mm]
			\hline
			&&&&\\[-4.0mm]
			pseudovector & $A_1$ & $A_2, A_3$ & $A_4, A_5$ & $A_6$ \\
			&&&&\\[-4.0mm]
			\hline
			&&&&\\[-4.0mm]
			tensor & $T_1$ & $T_2, T_3, T_7$ & $T_4, T_5, T_8$ & $T_6$ \\
			&&&&\\[-4.0mm]
			\hline
			&&&&\\[-4.0mm]
			scalar & & $S_1$ & $S_2$ & \\
			&&&&\\[-4.0mm]
			\hline
			&&&&\\[-4.0mm]
			pseudoscalar & & $P_1$ & $P_2$ & \\[1mm]
			\hline
		\end{tabular}
	\end{center}
\end{table} \noindent
Based on these considerations, we can order the calligraphic quantities in Eq. \eqref{eq:DecomNuclDAs} into the different twist contributions. 

Thus, we the calligraphic quantities contributing to twist four are:
\begin{table}[H]
	\begin{center}
		\begin{tabular}{|c||c|c|}
			\hline
			&&\\[-4.0mm]
			$\mathcal{F}$  & Integrand on r.h.s. of (\ref{eq:CalRelation}) & Abbreviation\\
			&&\\[-4.0mm]
			\hline
			\hline 
			&&\\[-4.0mm]
			${\cal S}_1$ & $S_1$ &  \\
			&&\\[-4.0mm]
			\hline 
			&&\\[-4.0mm]
			${\cal P}_1$ & $P_1$ & \\
			&&\\[-4.0mm]
			\hline 
			&&\\[-4.0mm]
			$2{\cal V}_3$ & $V_3$ & \\
			&&\\[-4.0mm]
			\hline 
			&&\\[-4.0mm]
			$2{\cal A}_3$ & $A_3$ & \\
			&&\\[-4.0mm]
			\hline 
			&&\\[-4.0mm]
			$2{\cal T}_3$ & $T_7$ & \\
			&&\\[-4.0mm]
			\hline 
			&&\\[-4.0mm]
			$2(P \cdot x){\cal V}_2$ & $V_1 - V_2 - V_3$ & $V_{123}$\\
			&&\\[-4.0mm]
			\hline 
			&&\\[-4.0mm]
			$2(P \cdot x){\cal A}_2$ & $- A_1 + A_2 - A_3$ & $A_{123}$\\
			&&\\[-4.0mm]
			\hline 
			&&\\[-4.0mm]
			$2(P \cdot x){\cal T}_2$ & $T_1 + T_2$ - $2 T_3$ & $T_{123}$ \\
			&&\\[-4.0mm]
			\hline 
			&&\\[-4.0mm]
			$2(P \cdot x){\cal T}_4$ & $T_1 - T_2 - 2 T_7$ & $T_{127}$\\[1mm]
			\hline
		\end{tabular}
	\end{center}
\end{table} \noindent
For brevity, the renormalization scale dependence is dropped in the following discussion. Moreover, we follow the notation from \cite{Braun:2000kw,Braun:2001tj,Braun:2006hz,Khodjamirian:2011jp}. The twist four DAs in the conformal expansion are given by:
\begin{align}
V_2(\alpha_i)=& \; 24 \alpha_1 \alpha_2 [\phi_4^0 + \phi_4^{+} (1- 5 \alpha_3)]\,,~~
A_2(\alpha_i) = 24 \alpha_1 \alpha_2 (\alpha_2 -\alpha_1) \phi_4^{-}\,, \nonumber\\
T_2(\alpha_i) =& \;24 \alpha_1 \alpha_2 [ \xi_4^{0} + \xi_4^{+} (1- 5 \alpha_3) ]\,, \nonumber\\
A_3(\alpha_i)=& \; 12 \alpha_3 (\alpha_2 - \alpha_1) [ (\psi_4^0+\psi_4^+) +  \psi_4^{-} (1-2 \alpha_3)]\nonumber \\
V_3(\alpha_i)=& \; 12 \alpha_3 [\psi_4^0 (1-\alpha_3) + \psi_4^{+}(1-\alpha_3 -10 \alpha_1 \alpha_2)+ \psi_4^{-}(\alpha_1^2 + \alpha_2^2 \nonumber \\ &- \alpha_3 (1 - \alpha_3)) ] \nonumber\\
T_3(\alpha_i)=& \; 6 \alpha_3  [(\phi_4^0 + \psi_4^0 + \xi_4^0 )(1-\alpha_3)
+(\phi_4^+ + \psi_4^+ + \xi_4^+ ) (1-\alpha_3 -10 \alpha_1 \alpha_2) \nonumber\\& 
+ (\phi_4^- - \psi_4^-+ \xi_4^- ) (\alpha_1^2 + \alpha_2^2 - \alpha_3 (1-\alpha_3))  ] \nonumber\\
T_7(\alpha_i)=& \; 6 \alpha_3 [ (\phi_4^0 + \psi_4^0 - \xi_4^0 )(1-\alpha_3)
+(\phi_4^+ + \psi_4^+ - \xi_4^+ ) (1-\alpha_3 -10 \alpha_1 \alpha_2) ] \nonumber\\&
+ (\phi_4^- - \psi_4^- - \xi_4^- ) (\alpha_1^2 + \alpha_2^2 - \alpha_3 (1-\alpha_3))  ] \nonumber\\
S_1(\alpha_i)=& \; 6 \alpha_3 (\alpha_2 -\alpha_1) [( \phi_4^0 + \psi_4^0 + \xi_4^0 +\phi_4^+ + \psi_4^+ + \xi_4^+ )
+ (\phi_4^- - \psi_4^- + \xi_4^- ) \nonumber \\ &\times (1- 2 \alpha_3)] \nonumber\\
P_1(\alpha_i)=& \; 6 \alpha_3 (\alpha_1 - \alpha_2) [( \phi_4^0 + \psi_4^0 - \xi_4^0 +\phi_4^+ + \psi_4^+ - \xi_4^+ )
+ (\phi_4^- - \psi_4^- - \xi_4^- ) \nonumber \\ &\times (1- 2 \alpha_3)]\,,
\label{eq:tw4DAs}
\end{align}
where we introduce additional parameters:
\begin{align}
	\phi_4^0 =& \; \frac12 (f_N + \lambda_1); \;  \phi_4^+ = \frac14 (f_N(3 - 10V_1^d) + \lambda_1(3 - 10 f_1^d)); \; \nonumber \\ \phi_4^- =& -\frac54 (f_N(1 - 2A_1^u) - \lambda_1(1 - 2 f_1^d - 4 f_1^u))
\end{align}
\begin{align}
	\psi_4^0 =& \; \frac12 (f_N - \lambda_1); \;  \psi_4^+ = - \frac14 (f_N(2 + 5A_1^u - 5V_1^d) - \lambda_1(2 - 5 f_1^d - 5 f_1^u)); \; \nonumber \\ \psi_4^- =& \; \frac54 (f_N(2 - A_1^u - 3 V_1^d) - \lambda_1(2 - 7 f_1^d + f_1^u))
\end{align}
\begin{align}
	\xi_4^0 =& \; \frac16 \lambda_2; \;  \xi_4^+ = \; \frac{1}{16} \lambda_2(4 - 15 f_2^d); \; \nonumber \\ \xi_4^- =& \; \frac{5}{16} \lambda_2(4 - 15 f_2^d)
\end{align}
For the purpose of a coherent discussion, we group them based on whether they belong to the leading or next-to-leading conformal spin \cite{Lenz:2009ar}:
\begin{table}[H]
	\begin{center}
		\begin{tabular}{|c||c|c|}
			\hline
			&&\\[-4.0mm]
			 & Leading twist & Higher twist \\
			&&\\[-4.0mm]
			\hline
			\hline
			&&\\[-4.0mm]
			Leading conformal spin & $f_N$ & $\lambda_1, \lambda_2$ \\
			&&\\[-4.0mm]
			\hline
			&&\\[-4.0mm]
			Next-to-leading conformal spin & $A_1^u, V_1^d$ & $f_1^u, f_1^d, f_2^d$ \\[1mm]
			\hline
		\end{tabular}
	\end{center}
\end{table} \noindent
For twist five, we have
\begin{table}[h]
	\begin{center}
		\begin{tabular}{|c||c|c|}
			\hline
			&&\\[-4.0mm]
			$\mathcal{F}$  & Integrand on r.h.s. of (\ref{eq:CalRelation}) & Abbreviation\\
			&&\\[-4.0mm]
			\hline 
			\hline
			&&\\[-4.0mm]
			$4 (P \cdot x){\cal V}_5$ & $V_4 - V_3$ & $V_{43}$ \\
			&&\\[-4.0mm]
			\hline 
			&&\\[-4.0mm]
			$4 (P \cdot x){\cal A}_5$ & $A_3 - A_4$ & $A_{34}$ \\
			&&\\[-4.0mm]
			\hline 
			&&\\[-4.0mm]
			$2 (P \cdot x){\cal T}_5$ & $-T_1 + T_5 + 2 T_8$ & $T_{158}$\\
			&&\\[-4.0mm]
			\hline 
			&&\\[-4.0mm]
			$4 (P \cdot x){\cal T}_7$ & $T_7 - T_8$ & $T_{78}$ \\
			&&\\[-4.0mm]
			\hline 
			&&\\[-4.0mm]
			$2 (P \cdot x){\cal S}_2$ & $S_1 - S_2$ & $S_{12}$\\
			&&\\[-4.0mm]
			\hline 
			&&\\[-4.0mm]
			$2(P \cdot x){\cal P}_2$ & $P_2 - P_1$ & $P_{21}$\\
			&&\\[-4.0mm]
			\hline 
			&&\\[-4.0mm]
			$4(P \cdot x){\cal V}_4$ & $- 2V_1 + V_3 + V_4 + 2V_5$ & $V_{1345}$\\
			&&\\[-4.0mm]
			\hline 
			&&\\[-4.0mm]
			$4(P \cdot x){\cal A}_4$ & $- 2A_1 - A_3 - A_4 + 2A_5$ & $A_{1345}$\\
			&&\\[-4.0mm]
			\hline 
			&&\\[-4.0mm]
			$4(P \cdot x)^2{\cal T}_6$ & $2T_2 - 2T_3 - 2 T_4 + 2T_5 + 2T_7 + 2T_8$ & $T_{234578}$\\
			&&\\[-4.0mm]
			\hline 
			&&\\[-4.0mm]
			${\cal V}_1^M$ & $V_1^M$ & \\
			&&\\[-4.0mm]
			\hline 
			&&\\[-4.0mm]
			${\cal A}_1^M$ & $A_1^M$ & \\
			&&\\[-4.0mm]
			\hline 
			&&\\[-4.0mm]
			${\cal T}_1^M$ & $T_1^M$ & \\[1mm]
			\hline
		\end{tabular}
	\end{center}
\end{table} 
\noindent
\begin{align}
V_4(\alpha_i) =& \; 3 [\psi_5^0 (1-\alpha_3) + \psi_5^{+} (1-  \alpha_3 - 2 (\alpha_1^2 +\alpha_2^2)) + \psi_5^{-} (2 \alpha_1 \alpha_2 \nonumber \\ &-\alpha_3 (1-\alpha_3)) ] \,,\nonumber\\
A_4(\alpha_i)=& \; 3  (\alpha_2 -\alpha_1) [ -\psi_5^{0} + \psi_5^{+}(1- 2 \alpha_3) + \psi_5^{-} \alpha_3] \,,\nonumber \\
T_4(\alpha_i)=& \;\frac{3}{2} [ (\phi_5^{0} + \psi_5^{0} + \xi_5^0 ) (1 - \alpha_3) + (\phi_5^{+} + \psi_5^{+} + \xi_5^{+} )  (1-  \alpha_3 - 2 (\alpha_1^2 + \alpha_2^2))]\nonumber\\
&+ (\phi_5^{-} - \psi_5^{-} + \xi_5^{-} )   (2 \alpha_1 \alpha_2 - \alpha_3 (1- \alpha_3)) \,, \nonumber\\
T_8(\alpha_i)=& \;\frac{3}{2}  [ ( \phi_5^0 + \psi_5^{0} - \xi_5^0 )
(1-\alpha_3) + (\phi_5^{+} +\psi_5^{+}-\xi_5^+ )(1-\alpha_3 -2 (\alpha_1^2 + \alpha_2^2) ) \nonumber\\
&+ (\phi_5^{-} - \psi_5^{-} + \xi_5^{-} )  (2 \alpha_1 \alpha_2 - \alpha_3 (1-
\alpha_3)) ]  \,, \nonumber \\
V_5(\alpha_i)=& \; 6 \alpha_3 [\phi_5^0 + \phi_5^+  (1-2 \alpha_3)  ]\,, ~~
A_5(\alpha_i)= 6 \alpha_3 (\alpha_2 - \alpha_1) \phi_5^- \,,\nonumber \\
T_5(\alpha_i)=& \;6 \alpha_3 [  \xi_5^0 + \xi_5^+  (1- 2 \alpha_3) ] \,, \nonumber\\
S_2(\alpha_i)=& \; \frac{3}{2} (\alpha_2 -\alpha_1) [-( \phi_5^0 + \psi_5^0 + \xi_5^0 )
+ (\phi_5^{+}+ \psi_5^{+} + \xi_5^{+} ) (1- 2 \alpha_3) \nonumber \\
&+ (\phi_5^{-} - \psi_5^{-} + \xi_5^{-} ) \alpha_3] \,, \nonumber \\
P_2(\alpha_i)=& \; \frac{3}{2} (\alpha_1 -\alpha_2) [-( \phi_5^0 + \psi_5^0 - \xi_5^0 )
+ (\phi_5^{+} + \psi_5^{+} - \xi_5^{+} ) (1- 2 \alpha_3)   \nonumber \\
&+ (\phi_5^{-} - \psi_5^{-} - \xi_5^{-} ) \alpha_3] \,, \nonumber
\end{align}
The new parameters in the twist five DAs can be expressed in the conformal expansion:
\begin{align}
	\phi_5^0 =& \; \frac12 (f_N + \lambda_1); \; \; \;  \phi_5^+ = - \frac{5}{6} [f_N  (3+4  V_1^d) - \lambda_1 (1- 4
	f_1^d)]\,,  \nonumber \\
	\phi_5^- =& - \frac{5}{3} [f_N  (1- 2 A_1^u) - \lambda_1 ( f_1^d - f_1^u)] 
\end{align}
\begin{align}
	\psi_5^0 =& \; \frac12 (f_N - \lambda_1); \;  \psi_5^+ = - \frac{5}{6} [f_N  (5+ 2 A_1^u - 2 V_1^d) -
	\lambda_1(1-2 f_1^d- 2 f_1^u)] \,,  \nonumber \\
	\psi_5^- =&  \frac{5}{3} [f_N  (2-  A_1^u - 3 V_1^d) + \lambda_1 ( f_1^d- f_1^u)]\,,
	\nonumber \\
\end{align}
\begin{align}
	\xi_5^0 = \frac16 \lambda_2; \; \; \; \xi_5^+ = \frac{5}{36} \lambda_2 (2 - 9 f_2^d); \; \; \,
	\xi_5^- = -\frac{5}{4} \lambda_2 f_2^d \,,
\end{align}
Finally, for twist six we obtain the following contributions:
\begin{table}[H]
	\begin{center}
		\begin{tabular}{|c||c|c|}
			\hline
			&&\\[-4.0mm]
			$\mathcal{F}$  & integrand on r.h.s. of (\ref{eq:CalRelation}) & Abbreviations \\
			&&\\[-4.0mm]
			\hline 
			\hline
			&&\\[-4.0mm]
			$4 (P \cdot x)^2{\cal V}_6$ & $-V_1 + V_2 + V_3 + V_4 + V_5 - V_6$ & $V_{123456}$ \\
			&&\\[-4.0mm]
			\hline 
			&&\\[-4.0mm]
			$4 (P \cdot x)^2{\cal A}_6$ & $A_1 - A_2 + A_3 + A_4 - A_5 + A_6$ & $A_{123456}$ \\
			&&\\[-4.0mm]
			\hline 
			&&\\[-4.0mm]
			$4 (P \cdot x)^2{\cal T}_8$ & $- T_1 + T_2 + T_5 - T_6 + 2T_7 + 2T_8$ & $T_{125678}$ \\[1mm]
			\hline
		\end{tabular}
	\end{center}
\end{table} \noindent
\begin{align}
V_6(\alpha_i)=& \;2 [  \phi_6^0  + \phi_6^+ (1- 3 \alpha_3) ] \,,~~
A_6(\alpha_i)= 2 (\alpha_2-\alpha_1)\phi_6^- \,,\nonumber \\
T_6(\alpha_i)=& \; 2 [ \phi_6^0  - {1 \over 2}(\phi_6^+-\phi_6^-) (1- 3 \alpha_3)]\,.
\end{align}
\noindent
The corresponding parameters read:
\begin{eqnarray}
	\phi_6^+ &=& \frac{1}{2} [f_N  (1-4  V_1^d) - \lambda_1 (1- 2 f_1^d)]\,,
	\nonumber \\
	\phi_6^- &=& \frac{1}{2} [f_N  (1+4  A_1^u) + \lambda_1 (1- 4 f_1^d - 2
	f_1^u)]\,.
\end{eqnarray}

\newpage
\clearpage
\section{Form factors}  \label{chp:AppendixFormFactors}

In this section, we state the remaining expressions for the form factors before the $z$-expansion.

\begin{align}
	\widetilde{F}^{(d)}_{B\to p_L}(q^2) =& \, \frac{1}{m_B^2 f_B} \int_0^{\alpha_0^B} \Md \alpha \, e^{\frac{m_B^2 - s(\alpha)}{M^2}} \Bigg\{\frac{m_b m_p}{4 \bar{\alpha}} \bigg(V_3(\alpha)-A_3(\alpha)\bigg)  + \frac{m_b^2 m_p^2}{2} \frac{\widetilde{S}_{12}(\alpha)-\widetilde{P}_{21}(\alpha)}{\bar{\alpha}^2M^2}  \nonumber \\ &+ \frac{m_b m_p}{4} \frac{\widetilde{A}_{123}(\alpha)-\widetilde{V}_{123}(\alpha)}{\bar{\alpha}^2} \bigg(1 - \frac{m_p^2 \bar{\alpha}^2 - q^2 + m_b^2}{\bar{\alpha} M^2}\bigg)  + \frac{m_b m_p^3}{2}  \bigg(1 + \frac{m_b^2}{\bar{\alpha}M^2}\bigg)\nonumber\\ &\times \frac{\widetilde{\widetilde{V}}_{123456}(\alpha)-\widetilde{\widetilde{A}}_{123456}(\alpha)}{\bar{\alpha}^2M^2} \Bigg\}
\label{eq:formdpl}
\end{align}
\begin{align}
	F^{(b)}_{B\to p_R}(q^2) =& \, \frac{1}{m_B^2 f_B} \int_0^{\alpha_0^B} \Md \alpha \, e^{\frac{m_B^2 - s(\alpha)}{M^2}} \Bigg\{\frac{m_b^2 m_p}{4} \bigg(\frac{m_p}{m_b} V_1(\alpha) - \frac{3}{\bar{\alpha}} T_1(\alpha) \bigg) \nonumber \\ &+ \frac{m_b^2 m_p}{4} \frac{P_1(\alpha)+S_1(\alpha)+6\cdot\widetilde{T}_7(\alpha)}{\bar{\alpha}} + \frac{m_b m_p^2}{2} \bigg(A_3(\alpha) - V_3(\alpha)\bigg)  \nonumber \\ & + \frac{m_b m_p^2}{4} \frac{\widetilde{V}_{123}(\alpha)-\widetilde{A}_{123}(\alpha)}{\bar{\alpha}} \bigg(1 + \frac{q^2 - m_p^2 \bar{\alpha}^2}{\bar{\alpha} M^2}\bigg) - \frac{m_b^2 m_p}{8} \frac{\widetilde{T}_{123}(\alpha)}{\bar{\alpha}^2} \nonumber \\ &\times \bigg(1 - \frac{m_b^2 - q^2 - m_p^2 \bar{\alpha}^2}{\bar{\alpha}M^2}\bigg) - \frac{3 m_b m_p^2}{8} \frac{\widetilde{A}_{34}(\alpha)+\widetilde{V}_{43}(\alpha)}{\bar{\alpha}} \bigg(1 + \frac{m_b^2}{\bar{\alpha} M^2}\bigg) \nonumber \\ &+ \frac{m_b^2 m_p^3}{4} \frac{\widetilde{P}_{21}(\alpha)-\widetilde{S}_{12}(\alpha)}{\bar{\alpha}M^2}  - \frac{3 m_b^2 m_p^3}{4} \frac{2\cdot\widetilde{T}_{78}(\alpha) +  \widetilde{T}_{158}(\alpha)}{\bar{\alpha}M^2}  \nonumber \\ &+ \frac{m_b m_p^2}{8} \frac{\widetilde{V}_{1345}(\alpha)+\widetilde{A}_{1345}(\alpha)}{\bar{\alpha}} \bigg(1 + \frac{m_b^2}{\bar{\alpha} M^2}\bigg)  + \frac{m_b m_p^4}{4 \bar{\alpha} M^2} \bigg(1 + \frac{m_b^2}{\bar{\alpha}M^2}\bigg) \bigg(\widetilde{A}_1^M-\widetilde{V}_1^M\bigg)(\alpha) \nonumber \\ & + \frac{3 m_b^4 m_p^3}{4} \frac{\widetilde{T}_1^M}{\bar{\alpha}^3 M^4} - \frac{m_b^2 m_p}{4} \frac{\widetilde{T}_{127}(\alpha)}{\bar{\alpha}}  \bigg(\frac{5}{2 \bar{\alpha}} \bigg(1 - \frac{m_b^2 - q^2}{\bar{\alpha}M^2}\bigg) - \frac{m_p^2}{2 M^2}\bigg) \nonumber \\ &+ \frac{3 m_b^4 m_p^3}{4} \frac{\widetilde{\widetilde{T}}_{125678}(\alpha)}{\bar{\alpha}^3 M^4} + \frac{m_b m_p^4}{4} \frac{\widetilde{\widetilde{A}}_{123456}(\alpha) - \widetilde{\widetilde{V}}_{123456}(\alpha)}{\bar{\alpha}M^2} \bigg(1 + \frac{m_b^2}{\bar{\alpha}M^2}\bigg) \nonumber \\ &- \frac{m_b^2 m_p^3}{8} \frac{\widetilde{\widetilde{T}}_{234578}(\alpha)}{\bar{\alpha}^2 M^2} \bigg(1 - \frac{m_p^2 \bar{\alpha}^2 - q^2 - m_b^2}{\bar{\alpha}M^2}\bigg) \Bigg\}\label{eq:formbpr}
\end{align}
\begin{align}
	\widetilde{F}^{(b)}_{B\to p_L}(q^2) =& \, \frac{1}{m_B^2 f_B} \int_0^{\alpha_0^B} \Md \alpha \, e^{\frac{m_B^2 - s(\alpha)}{M^2}} \Bigg\{\frac{m_b m_p^2}{4\overline{\alpha}} (V_1 (\alpha) + A_1 (\alpha)) + \frac{m_b m_p}{2 \bar{\alpha}} \left(A_3(\alpha)-V_3(\alpha)\right) \nonumber\\ &+ \frac{m_b^2 m_p^2}{4} \frac{\widetilde{P}_{21}(\alpha)-\widetilde{S}_{12}(\alpha)}{\bar{\alpha}^2M^2} + \frac{m_b m_p}{4\bar{\alpha}^2} \bigg(1 + \frac{q^2 - m_p^2 \bar{\alpha}^2 - m_b^2}{\bar{\alpha} M^2}\bigg) \Big(\widetilde{V}_{123}(\alpha) - \widetilde{A}_{123}(\alpha)\Big)  \nonumber \\ &- \frac{3 m_b^2 m_p^2}{4} \frac{2\cdot\widetilde{T}_{78}(\alpha)+\widetilde{T}_{158}(\alpha)}{\bar{\alpha}^2M^2} + \frac{m_b m_p^3}{4 \bar{\alpha}^2 M^2} \bigg(1 + \frac{m_b^2}{\bar{\alpha}M^2}\bigg) \left(\widetilde{A}_1^M(\alpha) -\widetilde{V}_1^M(\alpha)\right)\nonumber \\ &- \frac{m_b^2 m_p^2}{2} \frac{\widetilde{T}_{127}(\alpha)}{\bar{\alpha}^2M^2} + \frac{m_b m_p^3}{4} \frac{\widetilde{\widetilde{A}}_{123456}(\alpha)-\widetilde{\widetilde{V}}_{123456}(\alpha)}{\bar{\alpha}^2M^2} \bigg(1 + \frac{m_b^2}{\bar{\alpha}M^2}\bigg) \nonumber \\ &- \frac{m_b^2 m_p^2}{4} \frac{\widetilde{\widetilde{T}}_{234578}(\alpha)}{\bar{\alpha}^3 M^2} \bigg(1 - \frac{m_p^2 \bar{\alpha}^2 - q^2 + m_b^2}{2\bar{\alpha}M^2}\bigg) - \frac{m_b^2 m_p^2}{4M^2} \frac{\widetilde{T}_{123}(\alpha)}{\bar{\alpha}^2}\Bigg\}\label{eq:formbpl}
\end{align}

\newpage

\printbibliography
\end{document}